\documentclass[aps,pre,twocolumn]{revtex4-1}

\usepackage{amsmath}
\usepackage{amssymb}
\usepackage{color}

\usepackage{graphicx}

\DeclareMathAlphabet{\mathitbf}{OML}{cmm}{b}{it}

\newcommand{\dbar}{{\,\mathchar'26\mkern-12mu d}}

\newcommand{\sFrac}[2]{{\textstyle\frac{#1}{#2}}}
\newcommand{\tripleCdot}{\stackrel{\mbox{\bf\scriptsize .}}{:}}
\newcommand{\quadCdot}{\stackrel{\mbox{\small :}}{:}}
\begin{document}
\title{The micromechanics of nonlinear plastic modes}
%\author{ Luka Gartner${}^{1}$ and Edan Lerner${}^{1}$}
\author{Edan Lerner}
%\affiliation{${}^1$ Institute for Theoretical Physics, Institute of Physics, University of Amsterdam,Science Park 904, 1098 XH Amsterdam, The Netherlands}
\affiliation{Institute for Theoretical Physics, Institute of Physics, University of Amsterdam,Science Park 904, 1098 XH Amsterdam, The Netherlands}

\begin{abstract}

Nonlinear plastic modes (NPMs) are collective displacements that are indicative of imminent plastic instabilities in elastic solids. In this work we formulate the atomistic theory that describes the reversible evolution of NPMs and their associated stiffnesses under external deformations. The deformation-dynamics of NPMs is compared to those of the analogous observables derived from atomistic linear elastic theory, namely destabilizing eigenmodes of the dynamical matrix and their associated eigenvalues. The key result we present and explain is that the dynamics of NPMs and of destabilizing eigenmodes under external deformations follow different scaling laws with respect to the proximity to imminent instabilities. In particular, destabilizing modes vary with a singular rate, whereas NPMs' exhibit no such singularity. As a result, NPMs converge much earlier than destabilizing eigenmodes to their common final form at plastic instabilities. This dynamical difference between NPMs and linear destabilizing eigenmodes underlines the usefulness of NPMs for predicting the locus and geometry of plastic instabilities, compared to their linear-elastic counterparts. 

%Nonlinear plastic modes (NPMs) are collective displacements that are indicative of imminent plastic instabilities in elastic solids. In this work we formulate the atomistic theory that describes the reversible evolution of NPMs and their associated stiffnesses under external deformations. The deformation-dynamics of NPMs are compared to those of the analogous observables derived from atomistic linear elastic theory, namely destabilizing eigenmodes of the dynamical matrix and their associated eigenvalues. The key result we present is that NPMs converge much earlier than destabilizing eigenmodes to their common final form at plastic instabilities; this dynamical difference between NPMs and linear destabilizing eigenmodes underlines the usefulness of NPMs for predicting the loci and geometry of imminent plastic instabilities, compared to their linear-elastic counterparts. %We present scaling arguments that explain the observed differences in the deformation-dynamics of these two mode types.

\end{abstract}

\maketitle

\section{introduction}
When a disordered elastic solid is subjected to external deformation, particle-scale plastic instabilities are inevitably encountered \cite{exist}, each accompanied by a rearrangement of a small set of particles conventionally coined as a `shear transformation', and some degree of energy dissipation \cite{argon_st, falk_review, falk_langer_stz}. The occurrence rate, micromechanical consequences, and interactions between these instabilities determine the macroscopic rate of plastic deformation, which is a key rheological observable that controls important material properties such as toughness and elastic limit \cite{schuh_review_2007}.

\begin{figure*}[!ht]
\centering
\includegraphics[width = 0.93\textwidth]{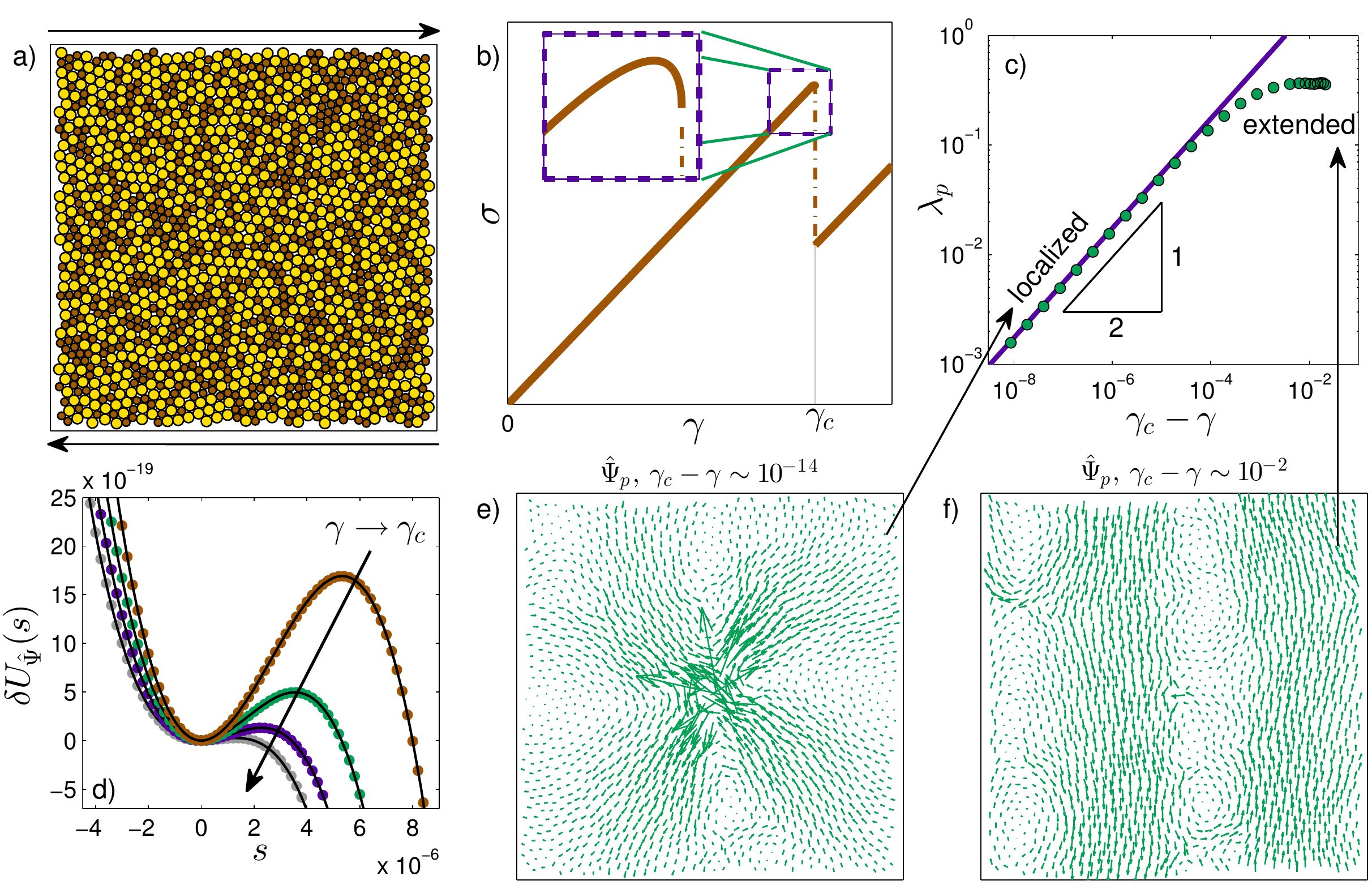}
\caption{\footnotesize Review of the micromechanics of a plastic instability. {\bf a)} An illustration of the basic setup considered in this work: an athermal glass under quasi-static simple shear deformation. {\bf b)} Cartoon of a typical stress $\sigma$ vs.~strain $\gamma$ signal in our setup; at some instability strain $\gamma_c$ a plastic instability occurs. The dashed frame shows that close to the instability the stress follows $\sigma - \sigma_c \sim \sqrt{\gamma_c - \gamma}$, as shown e.g.~in~\cite{lemaitre2004}. {\bf c)} Lowest eigenvalue $\lambda_p = {\cal M}:\hat{\Psi}_p\hat{\Psi}_p$ of the dynamical matrix ${\cal M}$, vs.~the distance in strain $\gamma_c - \gamma$ to the instability. Away from the instability the eigenmode $\hat{\Psi}_p$ associated with $\lambda_p$ is delocalized, as demonstrated in panel {\bf f)}, and $\lambda_p$ is largely insensitive to the deformation. As the solid is further deformed $\hat{\Psi}_p$ destabilizes and localizes, as demonstrated in panel {\bf e)}. $\lambda_p$ then vanishes as $\sqrt{\gamma_c - \gamma}$. {\bf d)} Energy variations $\delta U_{\hat{\Psi}}(s)$ upon displacing the particles about the mechanical equilibrium state according to $\delta\vec{x} = s\hat{\Psi}_p$, measured at distances $\gamma_c - \gamma = 10^{-14}, 10^{-41/3},10^{-40/3}$, and $10^{-13}$ away from the instability strain. These curves demonstrate the well-understood saddle-node bifurcation which characterizes plastic instabilities, in which a saddle point and minimum on the potential energy landscape coalesce and mutually anihilate, as shown in \cite{Malandro_Lacks,lemaitre2004,lemaitre2006_avalanches}. The continuous lines are obtained by a \emph{cubic} Taylor expansion of the energy variation, for which the expansion coefficients were calculated using inherent state information.}
\label{introduction_fig}
\end{figure*}

The micromechanical process that takes place as plastic instabilities are triggered under athermal conditions has been thoroughly studied in the framework of atomistic linear elasticity \cite{Malandro_Lacks,lemaitre2004,lemaitre2006_avalanches}. In this framework, plastic instabilities are reflected by the continuous vanishing of the lowest eigenvalue $\lambda_p$ of the dynamical matrix ${\cal M} \equiv \frac{\partial^2 U}{\partial \vec{x}\partial \vec{x}}$ (see Appendix for tensoric notation conventions) as the imposed shear strain $\gamma$ approaches an instability strain $\gamma_c$. Here and in what follows, $\vec{x}$ denotes the multidimensional coordinate vector of all particles' positions, and $U(\vec{x})$ denotes the potential energy. In the potential energy landscape (PEL) picture, plastic instabilities are understood as the coalescence and mutual annihilation of a local minimum and a nearby first-order saddle point, a process known as a saddle-node bifurcation, at some critical instability strain $\gamma_c$. This implies that asymptotically close to the instability strain, i.e.~as $\gamma \to \gamma_c$, the eigenvalue associated with the destabilizing eigenmode depends on the strain as $\lambda_p \sim \sqrt{\gamma_c - \gamma}$. In Fig.~\ref{introduction_fig} key micro and macroscopic aspects of the mechanics of plastic instabilities are reviewed.

Plastic instabilities are cleanly captured by destabilizing eigenmodes only very close (in strain) to instability strains, and more so as larger systems are considered, due to hybridization processes of destabilizing eigenmodes with low-energy plane waves. This is not the case, however, with \emph{nonlinear plastic modes} (NPMs), introduced first in \cite{luka}. NPMs are collective displacement directions which are indicative of the spatial structure and geometry of imminent plastic instabilities. Their definition, which is solely based on inherent structure information, hinges on properly accounting for the relevant anharmonicities of the potential energy landscape, as shown in \cite{luka} and explained in what follows. In this work we show that NPMs closely resemble plastic instabilities well away from instability strains, and well before destabilizing modes do. This is the case since NPMs do not `compete' with other low-frequency modes for their identity as the lowest-lying normal mode. They therefore do not suffer hybridizations with other modes, which leads to the preservation of their spatial structure remarkably far (in strain) from plastic instabilities. This superior robustness of NPMs identities renders their spatial distribution useful as means for a microstructural characterization of disordered solids that controls plastic deformation rates.

In this work we present a complete micromechanical theory for the deformation-dynamics of NPMs (i.e.~their evolution under imposed deformations) and their associated stiffnesses upon approaching plastic instabilities. The latter are compared to the deformation-dynamics of the conventional set of `linear' observables, namely the destabilizing eigenmodes of the dynamical matrix and their associated eigenvalues. In addition to demonstrating the persistence of NPM's identities over very large strain scales away from plastic instabilities, we further show that NPMs converge much faster \emph{scaling-wise} to their form at the instability strains, compared to destabilizing eigenmodes. We present a scaling analysis that explains the qualitative differences observed between the deformation-dynamics of these two types of modes. 

%%%%%%%%%%%%%%%%%%%%%%%%%%%%%%%%%%%%%%%%%%%%%%%%%%%%%%%
%%%%%%%%%%%%%%%%%%%%%%%%%%%%%%%%%%%%%%%%%%%%%%%%%%%%%%

This manuscript is organized as follows: in Sect.~\ref{numerics} we briefly describe the numerical methods and models used in this work. Further details about the numerics and algorithms used are provided in the Appendix, as are explanations of the tensor notations used throughout our work. In Sect.~\ref{linear_elasticity} we review the conventional micromechanical theory of plastic instabilities, discuss its range of applicability, and validate the theory against numerical simulations. In Sect.~\ref{plasic_modes} we reintroduce the barrier function, put forward first in \cite{luka}, from which the definition of NPMs emerges. We present results from a numerical investigation of the spatial properties of NPMs which are important for understanding NPMs deformation dynamics. We further present the micromechanical theory that describes the deformation dynamics of stiffnesses associated with NPMs. In Sect.~\ref{modes} we derive the micromechanical theory for the deformation dynamics of destabilizing modes and NPMs, and present data from numerical simulations that validate the theory's predictions. We end with a summary and discussion in Sect.~\ref{summary}.

\section{Methods and models}
\label{numerics}

We provide here a brief overview of the numerics used in our work; a complete and detailed description is provided in the Appendix. We employed a simple glass former in two dimensions that consists of point-like particles interacting via inverse power-law purely repulsive pairwise potentials. We expect our results to be independent of this particular choice of model. An example of a snapshot of our model glass with $N=1600$ is displayed in Fig.~\ref{introduction_fig}a.  We investigated systems of $N=40^2,80^2,160^2$ and $320^2$ particles; for each system size, we selected a single realization for which the first plastic instability upon shearing occured at a strain $\gamma_c \gtrsim 10^{-3}$. No other considerations were used when selecting each realization for the subsequent analyses carried out. All deformation simulations were carried out under athermal, quasi-static conditions, and the imposed deformation was simple shear under Lees-Edwards periodic boundary conditions. 128-bit numerics were employed to enable approaching plastic instabilities up to strains of the order of $\gamma_c -\gamma\sim 10^{-14}$. The calculation of nonlinear plastic modes (defined in Sect.~\ref{plasic_modes}) is explained in the Appendix.

\section{Plastic instabilities as reflected by atomistic linear elasticity}
\label{linear_elasticity}
In this Section we review the conventional atomistic theory of plastic instabilities in disordered elastic solids. The majority of the formalism presented in this Section appears in e.g.~\cite{lemaitre2004,lemaitre2006_avalanches,steady_states_with_jacques,barriers_lacks_maloney}; it is summarized here for the sake of completeness.

We consider a disordered system of $N$ particles in $\dbar$ dimensions, enclosed in a box of volume $\Omega$ under periodic boundary conditions, and interacting via some potential energy $U(\vec{x})$ which is a function of the particles' coordinates $\vec{x}$. Here and in all that follows, we restrict the discussion to the athermal limit $T \to 0$, with $T$ denoting the temperature. In the athermal limit, as long as it is mechanically stable, the system resides in a local minimum of the potential energy, i.e.~in a state $\vec{x}_0$ of mechanical equilibrium. This means that $(i)$ $\frac{\partial U}{\partial \vec{x}}\big|_{\vec{x}_0} = 0$ and $(ii)$ all eigenvalues of the dynamical matrix ${\cal M}\equiv \frac{\partial^2U}{\partial\vec{x}\partial\vec{x}}\big|_{\vec{x}_0}$ are non-negative (see Appendix for tensoric notation conventions). 

We next consider what happens when we deform our solid under quasi-static shear deformation, and in particular, we study how the eigenvalues of ${\cal M}$ vary as deformation is imposed. We start by writing the eigenmode decomposition of the dynamical matrix as
\begin{equation}
{\cal M} = \sum\limits_{\ell = 0}^{N\dbar}\lambda_{\ell} \hat{\Psi}_\ell\hat{\Psi}_\ell\,,
\end{equation}
where the orthonormal eigenmodes $\hat{\Psi}_\ell$ satisfy the eigenvalue equation
\begin{equation}
{\cal M}\cdot\hat{\Psi}_\ell = \lambda_\ell\hat{\Psi}_\ell\,,
\end{equation}
and therefore $\lambda_\ell = {\cal M}:\hat{\Psi}_\ell\hat{\Psi}_\ell$. We aim to spell out the deformation-dynamics of the eigenvalues, namely to derive an equation for $\frac{d\lambda_\ell}{d\gamma}$. In the athermal limit, total derivatives with respect to strain are taken according to
\cite{lutsko,lemaitre2004,lemaitre2006_avalanches,athermal_elasticity}
\begin{equation}\label{foo00}
\frac{d}{d\gamma} = \frac{\partial}{\partial \gamma} + \frac{d\vec{x}}{d\gamma}\cdot\frac{\partial}{\partial \vec{x}}\,,
\end{equation}
where $\frac{d\vec{x}}{d\gamma}$ denotes what are known as the \emph{nonaffine} part of the deformation-dynamics of the particles' coordinates. An explicit expression for $\frac{d\vec{x}}{d\gamma}$ can be derived by requiring that mechanical equilibrium is preserved under the deformation, namely 
\begin{equation}
\frac{d}{d\gamma}\frac{\partial U}{\partial \vec{x}} = \frac{\partial^2 U}{\partial \gamma \partial \vec{x}} + \frac{d\vec{x}}{d\gamma}\cdot\frac{\partial^2U}{\partial \vec{x}\partial \vec{x}} = 0\,,
\end{equation}
which can be inverted in favor of $\frac{d\vec{x}}{d\gamma}$, as
\begin{equation}\label{nonaffine}
\frac{d\vec{x}}{d\gamma} = -{\cal M}^{-1}\cdot\frac{\partial^2 U}{\partial \vec{x}\partial \gamma }\,.
\end{equation}
The superscript $-1$ should be understood here and in what follows as denoting the inverse of an operator taken after removing its zero modes. This removal is justified by the perfect orthogonality of the contracted vector with the zero modes of the inverted operator (which will always be the case in what follows). Eqs.~(\ref{foo00}) and (\ref{nonaffine}), introduced first in \cite{lutsko}, are central for the calculations presented in the subsequent sections. 

Using the formalism explained above, we take the derivative of an eigenvalue of ${\cal M}$ as
\begin{equation}\label{foo03}
\frac{d\lambda_\ell}{d\gamma} = \frac{d{\cal M}}{d\gamma}\!:\!\hat{\Psi}_\ell\hat{\Psi}_\ell = \frac{\partial {\cal M}}{\partial\gamma}\!:\!\hat{\Psi}_\ell\hat{\Psi}_\ell + U'''\!\tripleCdot\!\hat{\Psi}_\ell\hat{\Psi}_\ell\frac{d\vec{x}}{d\gamma}\,,
\end{equation}
where $U''' \equiv \frac{\partial^3U}{\partial\vec{x}\partial\vec{x}\partial\vec{x}}$, and no additional terms appear since normalization of modes implies that $\frac{d\hat{\Psi}_\ell}{d\gamma}\!\cdot\!\hat{\Psi}_\ell = 0$. Using the eigenmode decomposition of the dynamical matrix in Eq.~(\ref{nonaffine}), and inserting it in Eq.~(\ref{foo03}) we find
\begin{equation}\label{foo02}
\frac{d\lambda_\ell}{d\gamma} =  \frac{\partial {\cal M}}{\partial\gamma}\!:\!\hat{\Psi}_\ell\hat{\Psi}_\ell - \sum_m\frac{(U'''\!\tripleCdot\!\hat{\Psi}_\ell\hat{\Psi}_\ell\hat{\Psi}_m)(\hat{\Psi}_m\!\cdot\!\frac{\partial^2U}{\partial\vec{x}\partial\gamma})}{\lambda_m}.
\end{equation}

Eq.~(\ref{foo02}) describes the deformation dynamics of any of the $N\dbar$ eigenvalues $\lambda_\ell$ of ${\cal M}$. Here, we focus in particular on the equation for the lowest eigenvalue $\lambda_p$; as a plastic instability at a strain $\gamma_c$ is approached $\lambda_p \to 0$, and the RHS in the above equation is then dominated by the term in the sum pertaining to the destabilizing mode (an example of the latter can be seen in Fig.~\ref{introduction_fig}e). As $\gamma \to \gamma_c$, we can therefore approximate
\begin{equation}
\frac{d\lambda_p}{d\gamma}\bigg|_{\gamma \to \gamma_c^{-}} \simeq - \frac{\tau_p \nu_p}{\lambda_p}\,,
\end{equation}
where we have defined the \emph{asymmetry} of a mode $\hat{\Psi}_\ell$ as $\tau_\ell\equiv U'''\tripleCdot\hat{\Psi}_\ell\hat{\Psi}_\ell\hat{\Psi}_\ell$, and its shear-force coupling as $\nu_\ell \equiv \frac{\partial^2U}{\partial\gamma\partial\vec{x}}\cdot\hat{\Psi}_\ell$. This limiting differential equation, together with the boundary condition $\lambda_p(\gamma_c) = 0$, can be trivially solved for $\lambda_p$ as
\begin{equation}\label{foo01}
\lambda_p(\gamma \to \gamma_c^{-}) \simeq \sqrt{2\tau_p\nu_p}\sqrt{\gamma_c - \gamma}\,,
\end{equation}
where we have assumed that $\tau_p$ and $\nu_p$ are regular at $\gamma_c$. In Fig.~\ref{introduction_fig}c the scaling $\lambda_p \sim \sqrt{\gamma_c - \gamma}$ is confirmed by computer simulations.

Let us review two important consequences of Eq.~(\ref{foo01}), demonstrated in Fig.~\ref{introduction_fig}. First, on the macroscopic level, the shear stress and modulus also show signatures of plastic instabilities, that are derivable from Eq.~(\ref{foo01}); in the athermal limit, the shear modulus is given by \cite{lutsko,lemaitre2004,lemaitre2006_avalanches}
\begin{equation}
\mu = \frac{1}{\Omega}\left(\frac{\partial^2 U}{\partial \gamma^2} + \frac{\partial^2U}{\partial\gamma\partial\vec{x}}\cdot\frac{d\vec{x}}{d\gamma}\right)\,.
\end{equation}
As $\gamma \to \gamma_c$, $\lambda_p \to 0$, then $\frac{d\vec{x}}{d\gamma} = -\sum_\ell\frac{\nu_\ell}{\lambda_\ell}\hat{\Psi}_\ell \to -\frac{\nu_p}{\lambda_p}\hat{\Psi}_p$, and the shear modulus can be approximated as
\begin{equation}
\mu \simeq -\frac{\nu_p^2}{\lambda_p} \sim -(\gamma_c - \gamma)^{-\frac{1}{2}}\,.
\end{equation}
Consequently, the departure of the stress from its value $\sigma_c$ at the instability strain is expected to follow 
\begin{equation}
\sigma - \sigma_c \sim \sqrt{\gamma_c - \gamma}\,,
\end{equation}
as illustrated in the cartoon in Fig.~\ref{introduction_fig}b, and shown in e.g.~\cite{lemaitre2004}. 

Eq.~(\ref{foo01}) also leads to insights on the microscopic mechanics; we define $\delta U_{\hat{\Psi}}(s)$ as the variation of the potential energy upon displacing the particles about the inherent state $\vec{x}_0$ according to $\delta \vec{x} \equiv \vec{x} - \vec{x}_0 = s\hat{\Psi}_p$. For small distances $s$ we can expand $\delta U_{\hat{\Psi}}(s)$ as
\begin{equation}\label{foo15}
\delta U_{\hat{\Psi}}(s) \simeq \sFrac{1}{2}\lambda_p s^2 + \sFrac{1}{6}\tau_p s^3\,.
\end{equation}
Fig.~\ref{introduction_fig}d displays the energy variations $\delta U_{\hat{\Psi}}(s)$ obtained at various strains approaching a plastic instability strain $\gamma_c$. The softening of the stiffness $\lambda_p = \frac{d^2U}{ds^2}$ upon approaching the instability, as predicted by Eq.~(\ref{foo01}), is apparent, as is the decreasing of the saddle point. From Eq.~(\ref{foo15}) we deduce that the saddle point occurs at $s_\star = -2\sFrac{\lambda_p}{\tau_p}$, with a magnitude of $\delta U_{\hat{\Psi}}(s_\star) = \sFrac{2}{3}\sFrac{\lambda_p^3}{\tau_p^2} \sim (\gamma_c - \gamma)^{\frac{3}{2}}$ following Eq.~(\ref{foo01}), as shown in \cite{barriers_lacks_maloney,steady_states_with_jacques}.

%%%%%%%%%%%%%%%%%%%%%%%%%%%%%%%%%%%%%%%%%%%%%%%%%%%%%%%
\begin{figure}[!ht]
\centering
\includegraphics[width = 0.48\textwidth]{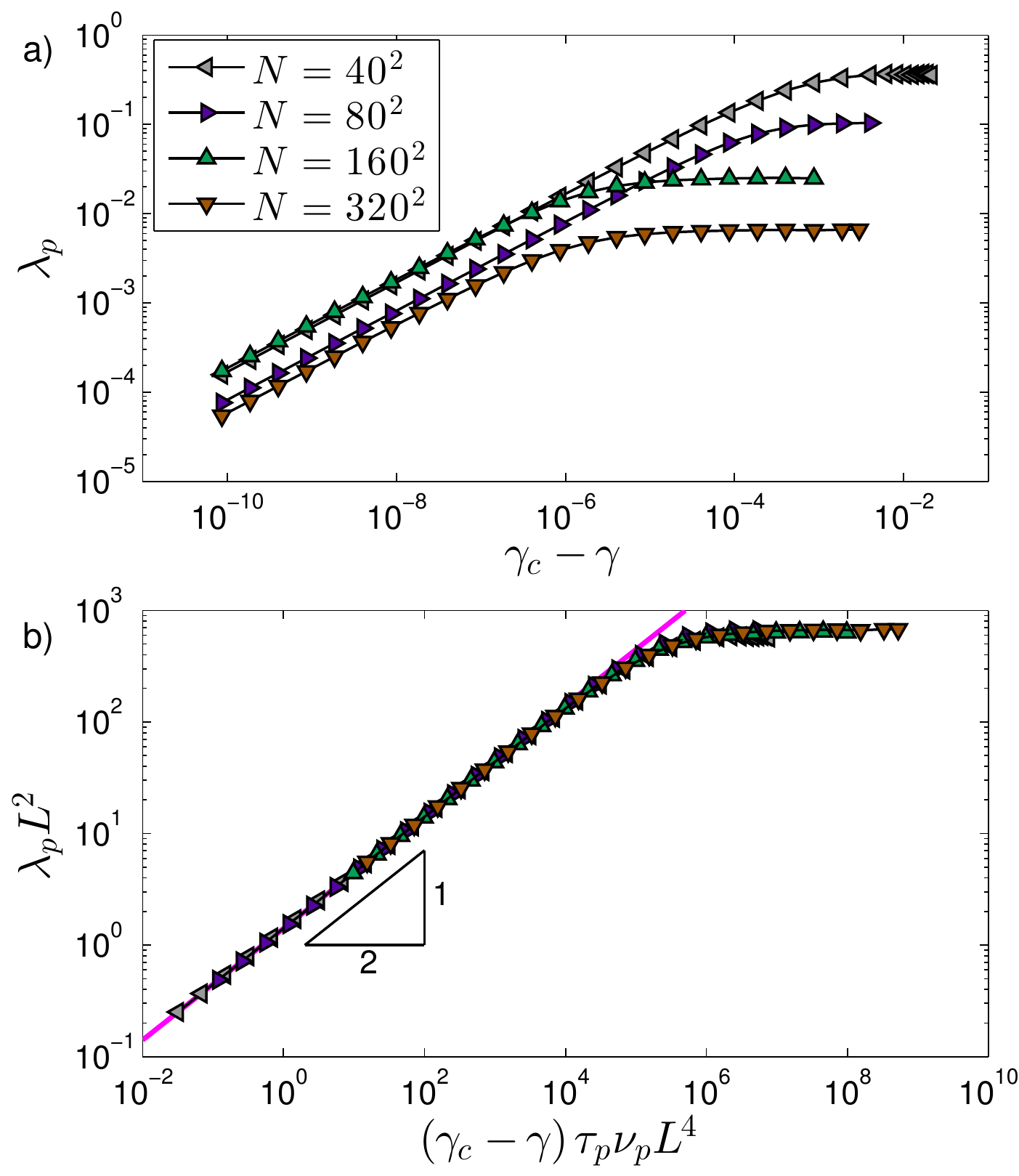}
\caption{\footnotesize {\bf a)} The eigenvalues $\lambda_p$ associated with destabilizing modes $\hat{\Psi}_p$ \emph{vs}.~distance in strain $\gamma_c - \gamma$ to a plastic instability strain $\gamma_c$, for various system sizes as shown in the legend. The instabilities were the first encountered upon shearing a randomly-selected freshly-quenched glass. {\bf b)} Rescaled $\lambda_p$'s vs.~the rescaled strain interval reveals that Eq.~(\ref{foo01}) holds on intervals below the scale $\delta\gamma \sim 1/(\tau_p\nu_pL^4)$. $\tau_p \equiv U'''\tripleCdot\hat{\Psi}_p\hat{\Psi}_p\hat{\Psi}_p$ and $\nu_p \equiv \frac{\partial^2 U}{\partial \gamma\partial\vec{x}}\cdot\hat{\Psi}_p$ were calculated at $\gamma_c - \gamma \lesssim 10^{-13}$.  }
\label{eigenvalue_fig}
\end{figure}
%%%%%%%%%%%%%%%%%%%%%%%%%%%%%%%%%%%%%%%%%%%%%%%%%%%%%%

How far away from the instability strain $\gamma_c$ is Eq.~(\ref{foo01}) valid? This depends on the strain scale in which the dehybridization of the destabilizing mode from the lowest plane-waves occurs, which can be estimated by comparing the stiffness associated with the lowest energy shear-wave in a system of linear size $L$, to the stiffness of the destabilizing mode $\lambda_p$. The former is expected to scale as $L^{-2}$, while the latter is proportional to $\sqrt{\tau_p\nu_p}\sqrt{\gamma_c - \gamma}$. Eq.~(\ref{foo01}) is therefore expected to hold at up to strain intervals $\gamma_c - \gamma \lesssim 1/(\tau_p\nu_pL^4)$, as indeed shown to hold numerically in Fig.~\ref{eigenvalue_fig}. In what follows we will show that this strain scale is central to the deformation dynamics of destabilizing modes.

\section{Nonlinear plastic modes (NPMs)}
\label{plasic_modes}
\subsection{Introduction and definitions}

The strain scale $1/(\tau_p\nu_pL^4)$ below which plastic instabilities are robustly reflected by the destabilizing mode quickly vanishes for large systems. An important question is therefore whether modes that are indicative of imminent plastic instabilities can be defined and detected away from instability strains, at scales $\gamma_c - \gamma \gg 1/(\tau_p\nu_pL^4)$. In other words, is it possible to overcome the difficulties associated with the hybridization of destabilizing modes with plane waves in the detection of imminent plastic instabilities. In \cite{luka} this question was answered to the affirmative: it was shown that nonlinear plastic modes (NPMs) exhibit remarkable resemblance to dehybridized destabilizing modes, and can be detected well before plastic instabilities, deep in the regime where the destabilizing mode is fully hybridized with plane waves. 

%We consider a system of $N$ particles in $\dbar$ dimensions, enclosed in a box of volume $\Omega$ under periodic boundary conditions, and interacting via some potential energy $U(\vec{x})$ which is a function of the particles' coordinates $\vec{x}$. Here and in all that follows, we restrict the discussion to the athermal limit $T \to 0$, with $T$ denoting the temperature. In the athermal limit, as long as it is mechanically stable, the system always resides in a local minimum of the potential energy, i.e.~it is in a state $\vec{x}_0$ of mechanical equilibrium. This means that $(i)$ $\frac{\partial U}{\partial \vec{x}}\big|_{\vec{x}_0} = 0$ and $(ii)$ all eigenvalues of the dynamical matrix ${\cal M}\equiv \frac{\partial^2U}{\partial\vec{x}\partial\vec{x}}\big|_{\vec{x}_0}$ are non-negative.

The theoretical framework within which the definition of NPMs emerges is constructed as follows: consider the variation $\delta U_{\hat{z}}(s)$ of the potential energy upon displacing the particles about the inherent state $\vec{x}_0$, but this time along a \textbf{general} collective displacement direction (mode) $\hat{z}$ (which may or may not be an eigenmode of ${\cal M}$), namely according to $\delta \vec{x} = s\hat{z}$. For small $s$, it writes
\begin{equation}\label{foo05}
\delta U_{\hat{z}}(s) \simeq \sFrac{1}{2}\kappa_{\hat{z}}s^2 + \sFrac{1}{6}\tau_{\hat{z}}s^3\,,
\end{equation}
where we have introduced the stiffness $\kappa_{\hat{z}} \equiv {\cal M}:\hat{z}\hat{z}$ and the asymmetry $\tau_{\hat{z}}\equiv U'''\tripleCdot\hat{z}\hat{z}\hat{z}$ associated with the collective displacement direction $\hat{z}$. Notice that the first order term in Eq.~(\ref{foo05}) is absent due to mechanical equilibrium, and $\hat{z}$ is dimensionless and normalized, i.e.~$\hat{z}\cdot\hat{z} = 1$. In its truncated form Eq.~(\ref{foo05}), $\delta U_{\hat{z}}$ possesses stationary points at $s=0$ and $s_\star = -\frac{2\kappa_{\hat{z}}}{\tau_{\hat{z}}}$, corresponding respectively to a minimum and maximum of the truncated potential energy variation along the reaction coordinate $s$. We emphasize that Eq.~(\ref{foo05}) differs from Eq.~(\ref{foo15}) by describing the energy variation upon displacing the particles along a \textbf{general} direction $\hat{z}$ in the former case, as oppose to along the eigenmode $\hat{\Psi}_p$ in the latter.

We next define the truncated energy variation at the maximum $s_\star$ as the `barrier function' $b(\hat{z})$, namely
\begin{equation}\label{foo06}
b(\hat{z}) \equiv \sFrac{1}{2}\kappa_{\hat{z}}s_\star^2 + \sFrac{1}{6}\tau_{\hat{z}}s_\star^3 =  \frac{2\kappa_{\hat{z}}^3}{3\tau_{\hat{z}}^2}\,.
\end{equation}
Notice that $b(\hat{z})$ is not a function of the reaction coordinate $s$, but instead a function of the multi-dimensional collective displacement direction $\hat{z}$. By construction, modes $\hat{z}$ for which $b(\hat{z})$ is small are characterized by small stiffnesses $\kappa_{\hat{z}}$ and large asymmetries $\tau_{\hat{z}}$. This, in turn, implies that the displacement distance $s_\star$ for those modes is small, and therefore the cubic expansion at distances $s \sim s_\star$ should be a faithful representation of the actual variation of the potential energy upon displacing the particles along $\hat{z}$, as demonstrated e.g.~for destabilizing modes in Fig.~\ref{introduction_fig}d. Thus, small enough $b$'s should pertain to actual saddle points (energy barriers) that separate between the inherent structure in which the system resides, and neighboring ones. 

NPMs are therefore defined as modes $\hat{\pi}$ for which $b$ attains a \emph{local minimum}. This means that modes $\hat{\pi}$ satisfy $\frac{\partial b}{\partial \vec{z}}\big|_{\vec{z} = \hat{\pi}} = 0$, and all eigenvalues of the linear operator $\frac{\partial^2b}{\partial\vec{z}\partial\vec{z}}\big|_{\vec{z} = \hat{\pi}}$ are non-negative. Local minima of $b$ do not guarantee the smallness of $b$, and therefore do not necessarily faithfully indicate an actual instability direction. Nevertheless, modes $\hat{\pi}$ that pertain to \emph{low-lying} minima of $b$ are indicative of directions that take the system over saddle points of the potential energy, and in particular indicate imminent plastic instabilities, as shown in \cite{luka}. 

At this point it is useful to note that the barrier function is invariant to variations of the norm of its vector argument, i.e.~$b(\hat{z}) = b(c\hat{z})$ for any finite $c$. This means that the barrier function can be equivalently expressed as a function of a set of $N\dbar$ \emph{independent} variables $\vec{z}$, as
\begin{equation}\label{barrier_function}
b(\vec{z}) = \frac{2}{3}\frac{\left( {\cal M}:\vec{z}\vec{z} \right)^3}{\left(U'''\tripleCdot\vec{z}\vec{z}\vec{z}\right)^2}\,.
\end{equation}
In turn, this allows us to meaningfully take partial derivatives with respect to those variables, and in particular
\begin{equation}\label{foo07}
\frac{\partial b}{\partial \vec{z}} = 4\frac{\kappa_{\vec{z}}^2}{\tau_{\vec{z}}^2}\left( {\cal M}\cdot\vec{z} - \frac{\kappa_{\vec{z}}}{\tau_{\vec{z}}}U''':\vec{z}\vec{z}\right)\,.
\end{equation}
The gradient $\frac{\partial b}{\partial \vec{z}}$ with respect to $\vec{z}$ given above vanishes when evaluated at NPMs $\hat{\pi}$, the latter are therefore solutions to the nonlinear equation
\begin{equation}\label{nonlinear_equation}
{\cal M}\cdot\hat{\pi} = \frac{\kappa_{\hat{\pi}}}{\tau_{\hat{\pi}}}U''':\hat{\pi}\hat{\pi}\,.
\end{equation}

Eq.~(\ref{nonlinear_equation}) is key to the deformation-dynamics of NPMs, and has an interesting geometric interpretation; to see this, imagine we displace the constituent particles about the inherent structure configuration according to $\delta\vec{x} = s\hat{\pi}$. The quadratic expansion in $s$ of the response force that results from this displacement is
\begin{equation}
\vec{F}_{\hat{\pi}}(s) \simeq -{\cal M}\cdot\hat{\pi}\,s - \sFrac{1}{2}U''':\hat{\pi}\hat{\pi}\,s^2\,.
\end{equation}
Eq.~(\ref{nonlinear_equation}) tells us that the linear and nonlinear coefficients of the force response expansion are \emph{parallel} $N\dbar$ dimensional vectors. %, and should therefore have the same spatial structure.  
%Eq.~(\ref{nonlinear_equation}) tells us that the linear and nonlinear terms of the force response ${\cal M}\cdot\hat{\pi}$ and $U''':\hat{\pi}\hat{\pi}$ respectively, are \emph{parallel} $N\dbar$ dimensional vectors. %, and should therefore have the same spatial structure.  

%%%%%%%%%%%%%%%%%%%%%%%%%%%%%%%%%%%%%%%%%%%%%%%%%%%%%%%
\begin{figure}[!ht]
\centering
\includegraphics[width = 0.5\textwidth]{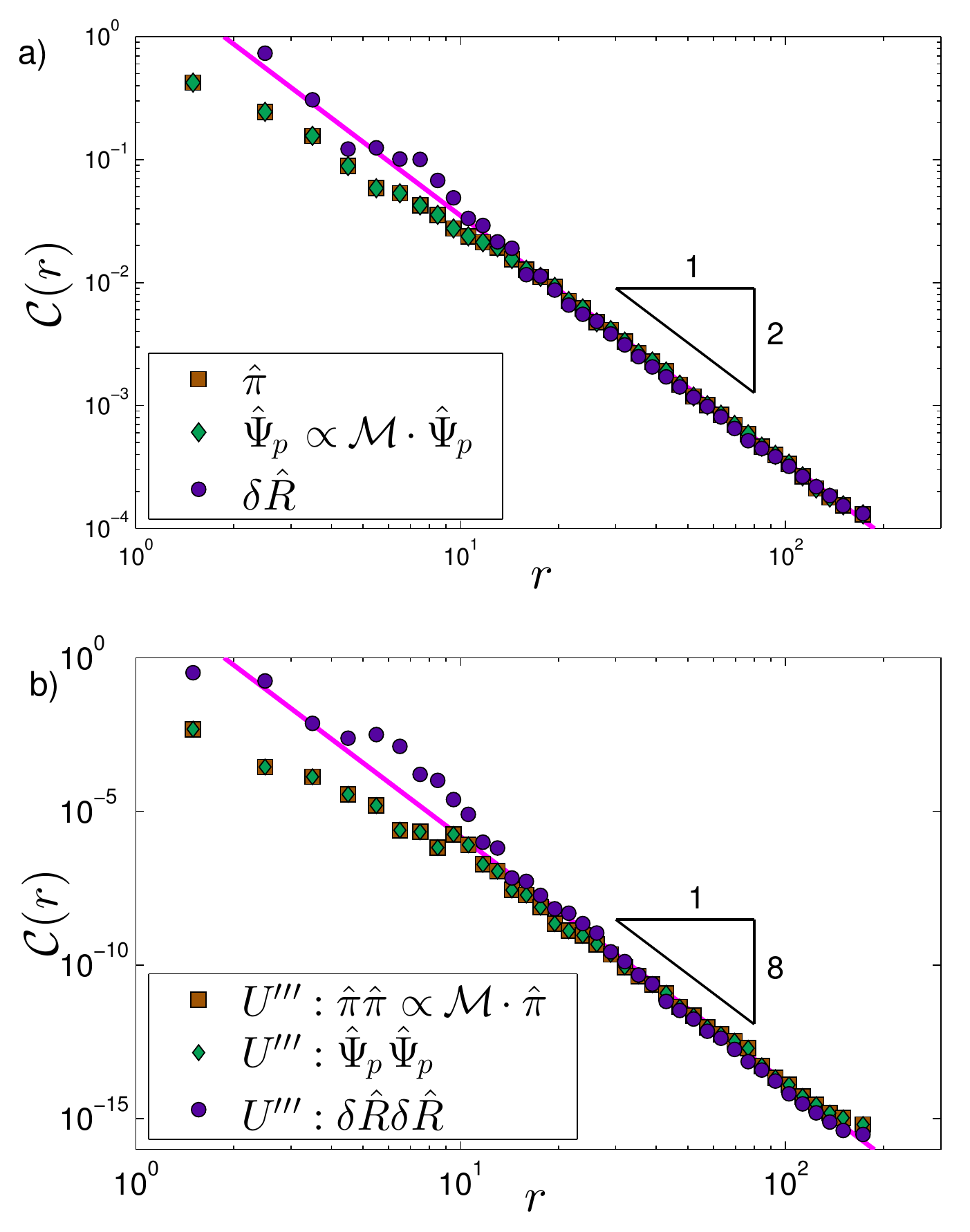}
\caption{\footnotesize {\bf a)} Spatial decay profiles ${\cal C}(r)$ (see text for definition) of a NPM $\hat{\pi}$ and of a destabilizing eigenmode $\hat{\Psi}_p$ calculated in a system of $N=102400$ at a distance $\gamma_c - \gamma \sim 10^{-14}$ away from a plastic instability. Also plotted is the decay profile of the response $\delta\vec{R} = {\cal M}^{-1}\vec{d}$ to a local dipolar force $\vec{d}$. All these modes are found to decay as $r^{-1}$ (notice that ${\cal C}$ scales as the magnitude squared of a mode's components). {\bf b)} The nonlinear force responses of the same modes analyzed in panel a) decay as $r^{-4}$. We verify that the double contraction of a spatially-decaying mode with $U'''$, e.g.~$U''':\hat{\Psi}_p\hat{\Psi}_p$, picks up the square of the spatial gradient of that mode, in consistency with Eq.~(\ref{foo08}). Notice that the linear and nonlinear force responses of the NPM $\hat{\pi}$ are parallel, therefore $|{\cal M}\cdot\hat{\pi}|(r) \sim r^{-4}$ as well, while $|{\cal M}\cdot\hat{\Psi}|(r) \sim r^{-1}$.} %This implies that the operation of ${\cal M}$ on $\hat{\pi}$ leads to a finite rotation of $\hat{\pi}$, see the discussion surrounding Eq.~(\ref{inequality}), and Fig.~\ref{pi_rotation_fig} below.}
\label{decay_profiles_fig}
\end{figure}
%%%%%%%%%%%%%%%%%%%%%%%%%%%%%%%%%%%%%%%%%%%%%%%%%%%%%%

What is the spatial structure of the said force response, and in particular of the parallel vectors ${\cal M}\cdot\hat{\pi}$ and $U''':\hat{\pi}\hat{\pi}$? In \cite{luka} it was shown that NPM's structure consists of a disordered, localized core, decorated by long-range largely-affine displacement fields that decay away from the core as $|\hat{\pi}|(r) \sim r^{1-\dbar}$, where $r$ denotes the distance from the NPM's core center. The force response $\vec{F}_{\hat{\pi}}$ is given by a double contraction of $\hat{\pi}$ with the third order tensor $U'''$; we therefore expect the relative magnitude of the force response away from the NPM's core to scale as the gradient squared of $\hat{\pi}$, namely 
\begin{equation}\label{foo08}
|\vec{F}_{\hat{\pi}}|(r) \sim |\nabla \hat{\pi}|^2(r) \sim r^{-2\dbar}\,.
\end{equation}
This relation is further motivated in the Appendix, for the simple case of pairwise central-force potentials.

To verify Eq.~(\ref{foo08}) numerically, we define the spatial decay profiles ${\cal C}_{\vec{v}}(r)$ which are calculated on a vector $\vec{v}$ by taking the median over the square of all components of the normalized $\hat{v}$ that are situated at a distance $\approx r$ away from the core of a plastic instability, see \cite{luka} for further details. In Fig.~\ref{decay_profiles_fig}a we plot the decay profiles of a NPM $\hat{\pi}$ (calculated as explained in the Appendix) and a destabilizing mode $\hat{\Psi}_p$ measured close to a plastic instability. These decay profiles are compared to that calculated for the displacement response $\delta\vec{R} = {\cal M}^{-1}\vec{d}$ to a local dipolar force $\vec{d}$ (as described in e.g.~\cite{breakdown}) in an undeformed solid. All three modes are found to decay as $r^{-1}$ (in our two-dimensional simulations). In Fig.~\ref{decay_profiles_fig}b we plot the spatial decay profiles of the double contractions of these three modes with the third-order tensor $U'''$. We indeed find that ${\cal C}_{\hat{\pi}}\sim r^{-2}$ and ${\cal C}_{{\cal M}\cdot\hat{\pi}} \sim {\cal C}_{U''':\hat{\pi}\hat{\pi}} \sim r^{-8}$ implying that $|\hat{\pi}|(r) \sim r^{-1}$, and $|{\cal M}\!\cdot\!\hat{\pi}|(r)  \sim |U'''\!:\!\hat{\pi}\hat{\pi}|(r) \sim r^{-4}$, supporting Eq.~(\ref{foo08}). 

The above discussion and the data plotted in Fig.~\ref{decay_profiles_fig} lead to an interesting conclusion: although destabilizing modes and NPMs share the same spatial decay profiles, the linear force responses ${\cal M}^{-1}\cdot\hat{\Psi}_p$ and ${\cal M}^{-1}\cdot\hat{\pi}$ do not; the former decay away from the disordered core as $r^{1-\dbar}$ (just as the destabilizing modes themselves), whereas the latter decay as $r^{-2\dbar}$.

\subsection{Dynamics of NPM stiffnesses}
\begin{figure}[!ht]
\centering
\includegraphics[width = 0.49\textwidth]{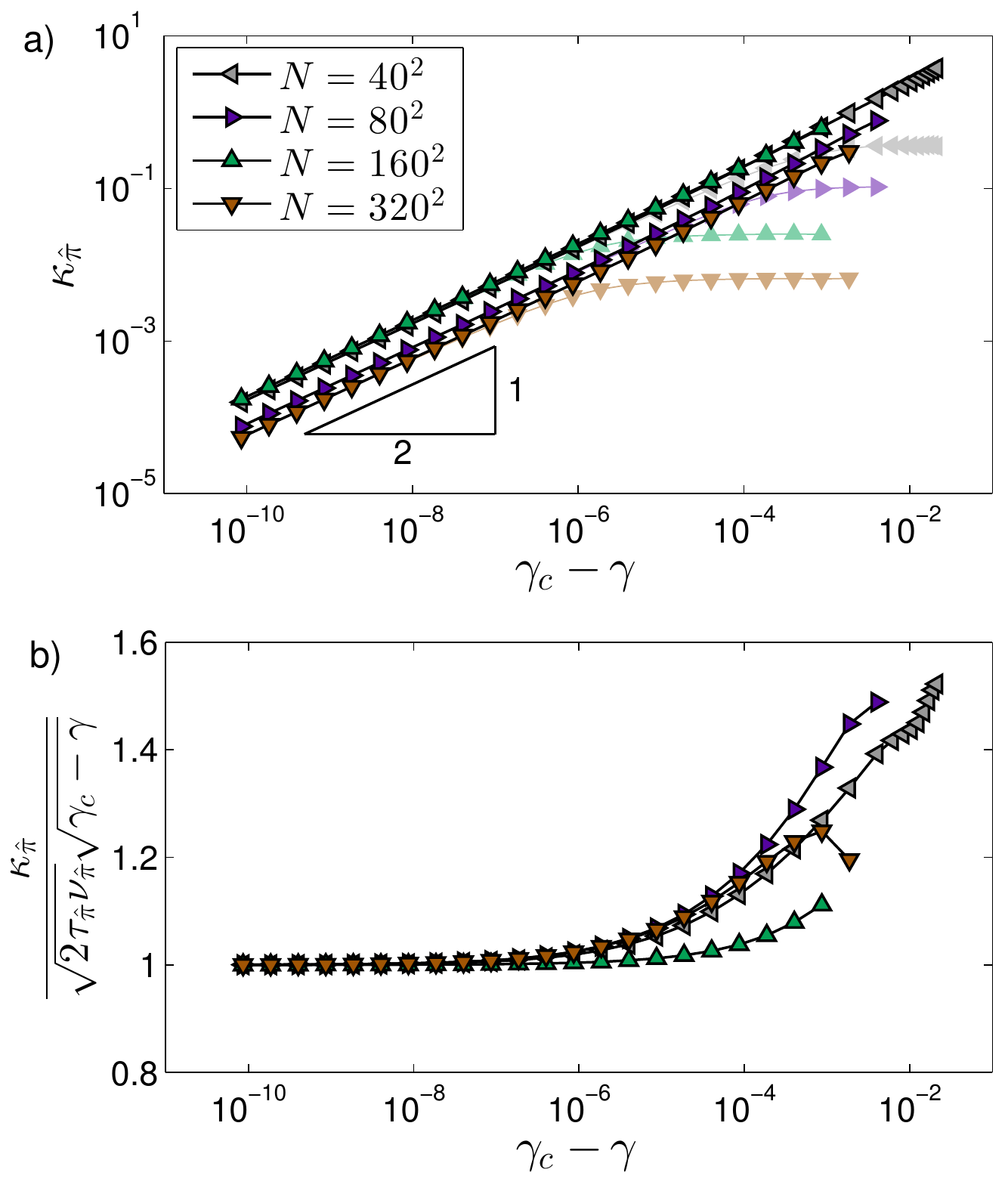}
\caption{\footnotesize {\bf a)} Stiffnesses $\kappa_{\hat{\pi}}$ \emph{vs}.~strain interval $\gamma_c - \gamma$. The pale symbols represent the eigenvalues $\lambda_p$ which perfectly coincide with the $\kappa_{\hat{\pi}}$'s as $\gamma \to \gamma_c$. {\bf b)} Rescaling the stiffnesses by their asymptotic form $\kappa_{\hat{\pi}}\simeq \sqrt{2\tau_{\hat{\pi}}\nu_{\hat{\pi}}}\sqrt{\gamma_c-\gamma}$ verifies Eq.~(\ref{kappa}), and shows that this scaling breaks down at a strain scale with no clear system-size dependence, in stark contrast with the eigenvalues $\lambda_p$ shown in Fig.~\ref{eigenvalue_fig}. Nevertheless, up to strain intervals $\gamma_c - \gamma\lesssim10^{-3}$, we find that the deviations from the asymptotic form remain less than roughly 50\%. }%{\bf c)} Shifting the rescaled stiffnesses by unity reveals that the subleading term of $\kappa_{\hat{\pi}}$ vs.~$\gamma_c - \gamma$ is linear in $\gamma_c- \gamma$. }
\label{kappa_fig}
\end{figure}

We next show that the deformation dynamics of NPMs stiffnesses $\kappa_{\hat{\pi}} = {\cal M}:\hat{\pi}\hat{\pi}$ and of the eigenvalues $\lambda_p = {\cal M}:\hat{\Psi}_p\hat{\Psi}_p$ obey the same equation of motion close to plastic instabilities. The total derivative with respect to deformation of the stiffness reads
\begin{eqnarray}
\frac{d\kappa_{\hat{\pi}}}{d\gamma} & = & \frac{d{\cal M}}{d\gamma}:\hat{\pi}\hat{\pi} + 2{\cal M}:\frac{d\hat{\pi}}{d\gamma}\hat{\pi} \nonumber \\
& = & U'''\tripleCdot\hat{\pi}\hat{\pi}\frac{d\vec{x}}{d\gamma} + \frac{\partial {\cal M}}{\partial \gamma}:\hat{\pi}\hat{\pi}+ 2{\cal M}:\frac{d\hat{\pi}}{d\gamma}\hat{\pi}\,. \nonumber 
\end{eqnarray}
Notice next that the first term on the RHS of the above equation can be written using Eqs.~(\ref{nonaffine}) and (\ref{nonlinear_equation}) as
\begin{equation}
U'''\tripleCdot\hat{\pi}\hat{\pi}\frac{d\vec{x}}{d\gamma} = -\frac{\tau_{\hat{\pi}}}{\kappa_{\hat{\pi}}}\hat{\pi}\cdot{\cal M}\cdot{\cal M}^{-1}\cdot\frac{\partial^2U}{\partial\vec{x}\partial\gamma} = -\frac{\tau_{\hat{\pi}} \nu_{\hat{\pi}}}{\kappa_{\hat{\pi}}}\,,
\end{equation}
and therefore we arrive at
\begin{equation}
\frac{d\kappa_{\hat{\pi}}}{d\gamma} = -\frac{\tau_{\hat{\pi}} \nu_{\hat{\pi}}}{\kappa_{\hat{\pi}}} + \frac{\partial {\cal M}}{\partial \gamma}:\hat{\pi}\hat{\pi}+ 2{\cal M}:\frac{d\hat{\pi}}{d\gamma}\hat{\pi}\,.
\end{equation}
The vanishing of $\kappa_{\hat{\pi}}$ upon plastic instabilities also implies that $|{\cal M}\cdot\hat{\pi}|\to 0$. Assuming that $|\frac{d\hat{\pi}}{d\gamma}| \lesssim |{\cal M}\cdot\hat{\pi}|^{-1}$ as plastic instabilities are approached (an assumption that will be established in the following Section), and recalling that $\frac{\partial {\cal M}}{\partial \gamma}$ is always regular, the last two terms in the RHS of the above equation can be neglected close to plastic instabilities, and we are left with
\begin{equation}
\frac{d\kappa_{\hat{\pi}}}{d\gamma}\bigg|_{\gamma \to \gamma_c^{-}} \simeq -\frac{\tau_{\hat{\pi}} \nu_{\hat{\pi}}}{\kappa_{\hat{\pi}}}\,.
\end{equation}
This limiting differential equation is identical in structure to Eq.~(\ref{foo01}) for the deformation dynamics of the eigenvalues $\lambda_p$ associated with destabilizing eigenmodes $\hat{\Psi}_p$. It is therefore solved by
\begin{equation}\label{kappa}
\kappa_{\hat{\pi}}(\gamma \to \gamma_c^{-}) \simeq \sqrt{2\tau_{\hat{\pi}} \nu_{\hat{\pi}}}\sqrt{\gamma_c - \gamma}\,,
\end{equation}
which is verified numerically in Fig.~\ref{kappa_fig}.

%%%%%%%%%%%%%%%%%%%%%%%%%%%%%%%%%%%%%%%%%%%%%%%%%%%%%%%
%%%%%%%%%%%%%%%%%%%%%%%%%%%%%%%%%%%%%%%%%%%%%%%%%%%%%%

One important observation to note is that Eq.~(\ref{kappa}) is followed over large strain intervals $\gamma_c - \gamma$, without a clear system-size dependence, as can be seen in Fig.~\ref{kappa_fig}. This stands in contrast with what is seen for the eigenvalues of destabilizing modes as described by Eq.~(\ref{foo01}), which is only valid over scales $\gamma_c - \gamma \lesssim L^{-4}$. This difference arises since NPMs do not `compete' for their identity with other low-frequency normal modes, i.e.~they do not suffer hybridizations. 

One obvious limitation on the range over which Eq.~(\ref{kappa}) is valid is the extent of typical elastic branches between consecutive plastic instabilities, which has been shown to vanish as $N^{-\beta}$ with $\beta \approx 2/3$ \cite{exist,my_yielding_rapid_pre_2010}. We therefore assert that above some system size the deformation dynamics of NPMs associated with imminent plastic instabilities will always be described by Eq.~(\ref{kappa}).

\begin{figure}[!ht]
\centering
\includegraphics[width = 0.48\textwidth]{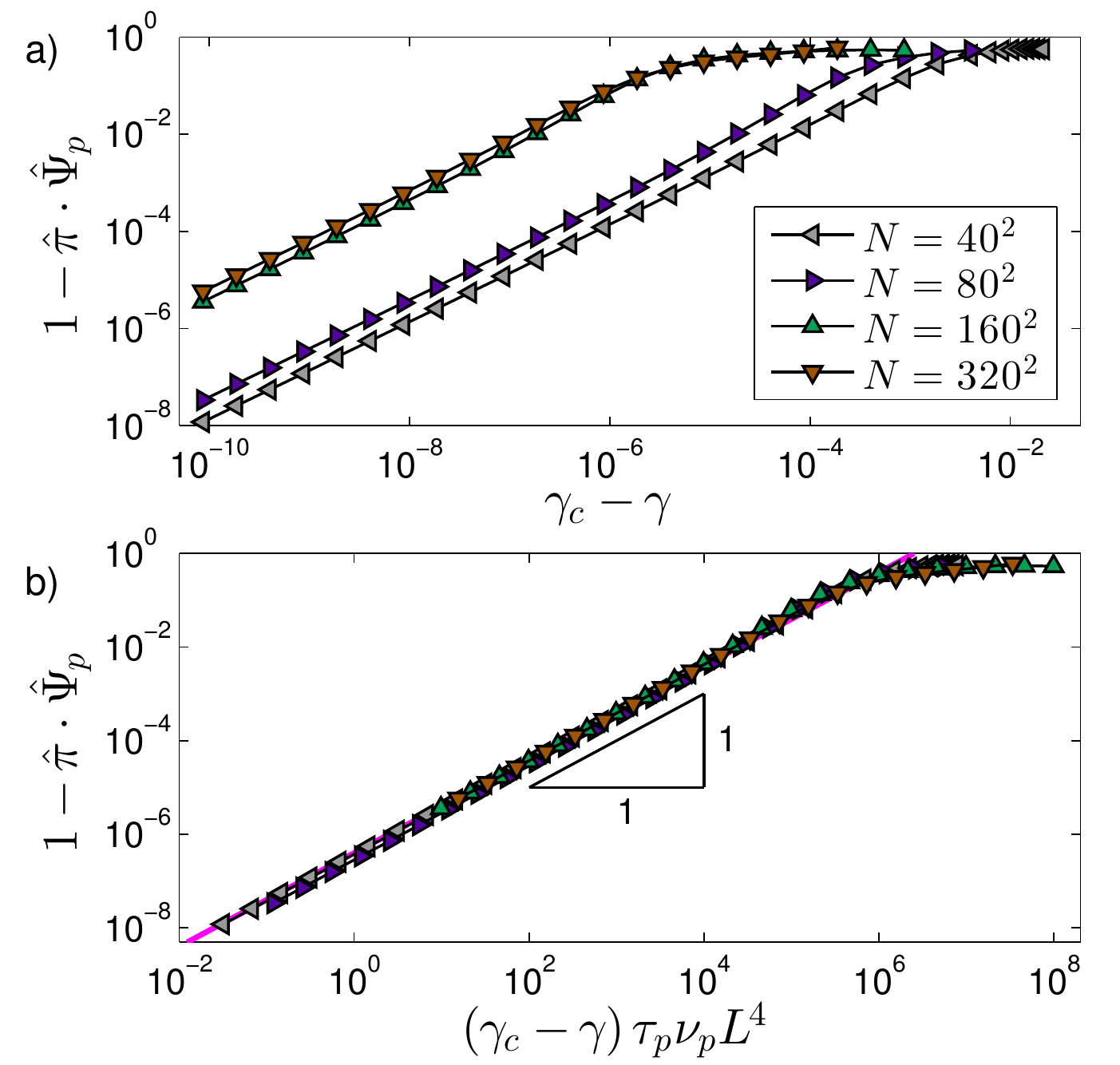}
\caption{\footnotesize {\bf a)} The NPM $\hat{\pi}$ and the destabilizing mode $\hat{\Psi}_p$ converge to a common final form at the instability strain $\gamma_c$, as indicated by the vanishing of $1-\hat{\pi}\cdot\hat{\Psi}_p$. {\bf b)} We find that $1-\hat{\pi}\cdot\hat{\Psi}_p \sim \gamma_c - \gamma$, see Sect.~\ref{modes} for a theoretical explanation of this scaling. We also find that the same strain scale $\delta\gamma \sim 1/(\tau_p\nu_pL^4)$ controls the convergence of both modes to their common final form.}
\label{overlap_pi_psi}
\end{figure}

Finally, we underline an important consequence of the simultaneous vanishing of the eigenmode $\lambda_p$ and the stiffness $\kappa$ at the same instability strain: both the destabilizing mode $\hat{\Psi}$ and the NPM $\hat{\pi}$ must converge to a common final form at the instability strain, since at that point they both satisfy ${\cal M}\cdot\hat{\Psi}_p  = {\cal M}\cdot\hat{\pi} = 0$ and must therefore be equal. This convergence of the two modes at a plastic instability is validated in Fig.~\ref{overlap_pi_psi}. 

%%%%%%%%%%%%%%%%%%%%%%%%%%%%%%%%%%%%%%%%%%%%%%%%%%%%%%%
%%%%%%%%%%%%%%%%%%%%%%%%%%%%%%%%%%%%%%%%%%%%%%%%%%%%%%

\section{Deformation dynamics of linear and nonlinear modes}
\label{modes}

\begin{figure}[!ht]
\centering
\includegraphics[width = 0.48\textwidth]{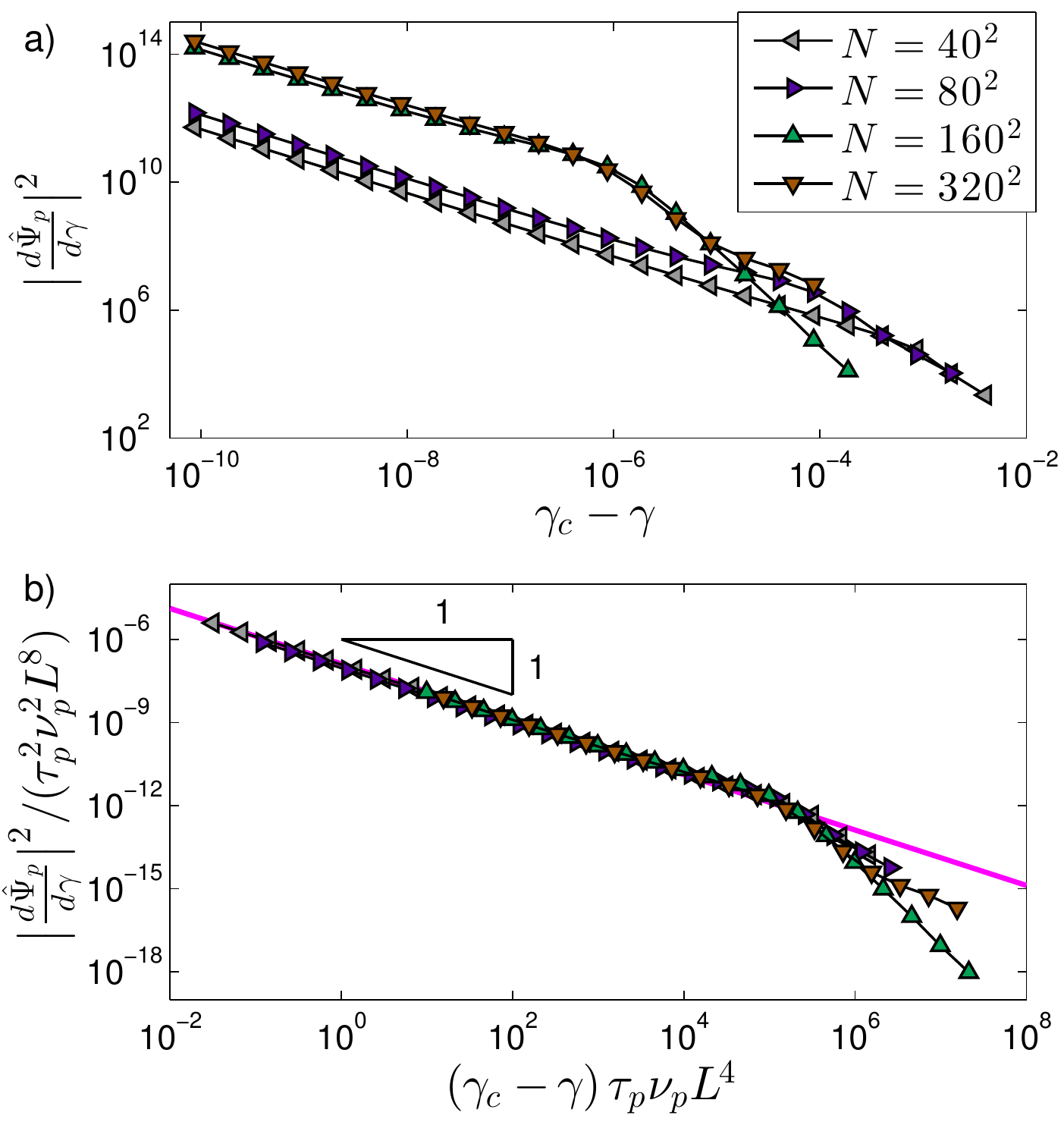}
\caption{\footnotesize {\bf a)} The total derivative squared with respect to strain of destabilizing eigenmodes, $\big| \frac{d\hat{\Psi}_p}{d\gamma}\big|^2$ \emph{vs.}~the distance to the imminent plastic instability strain $\gamma_c - \gamma$. {\bf b)} An appropriate rescaling (see text) reveals that the same strain scale $\delta\gamma \sim 1/(\tau_p\nu_pL^4)$ controls the deformation dynamics of destabilizing modes, as well as their associated eigenvalues.}
\label{eigenmode_fig}
\end{figure}

\begin{figure}[!ht]
\centering
\includegraphics[width = 0.43\textwidth]{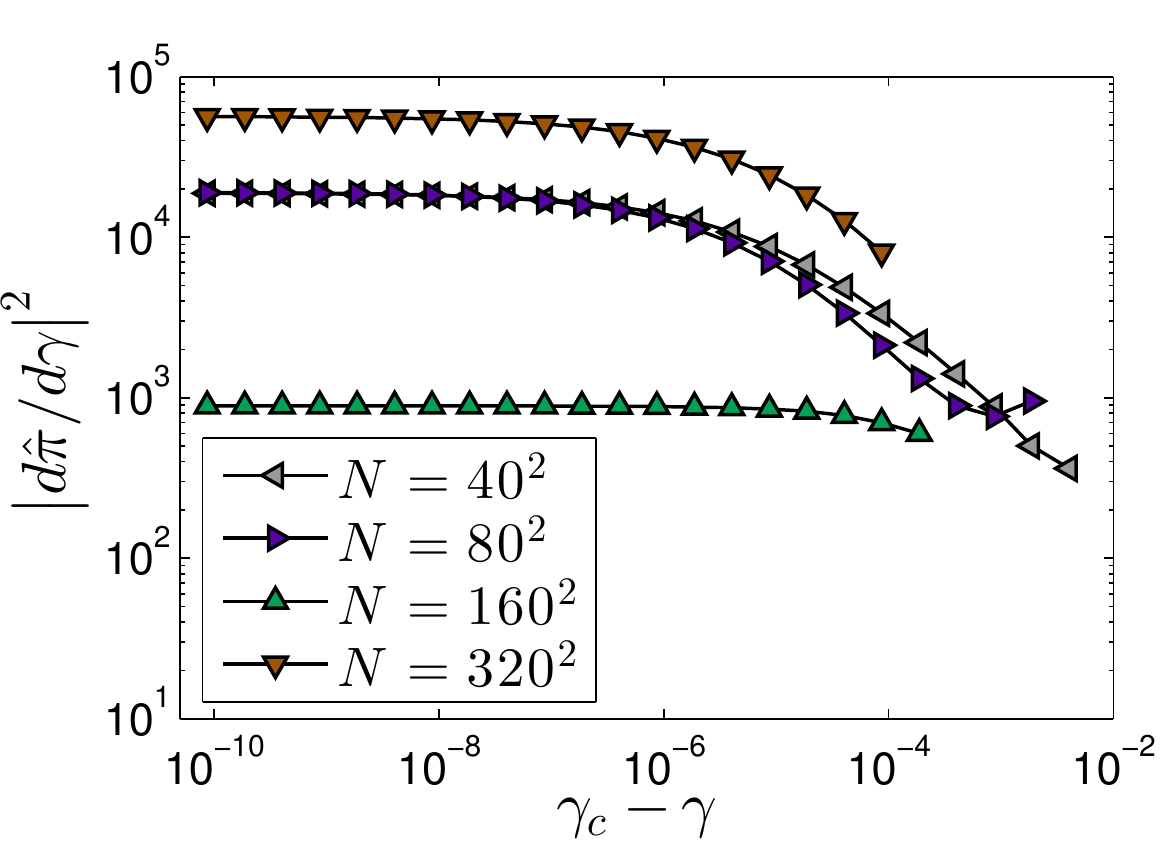}
\caption{\footnotesize Norm squared of the total derivatives of NPMs \emph{vs}.~$\gamma_c-\gamma$. Although NPM stiffnesses and eigenvalues associated with destabilizing modes follow the same scaling $\kappa_{\hat{\pi}}\sim \lambda_p \sim\sqrt{\gamma_c - \gamma}$, the two modes' deformation dynamics follow different scaling laws, namely $\big|d\hat{\Psi}_p/d\gamma\big|^2 \sim (\gamma_c - \gamma)^{-1}$, while $\big|d\hat{\pi}/d\gamma\big|^2 \sim (\gamma_c - \gamma)^0$. }
\label{pi_variation_fig}
\end{figure}

We have seen theoretically and numerically that the stiffnesses associated with destabilizing modes and NPMs are enslaved to the same equation of motion at scales $\gamma_c - \gamma \lesssim L^{-4}$ away from to plastic instabilities. Is there a similar equivalence between the deformation-dynamics of the destabilizing mode and that of the NPM? In this section we derive exact equations of motion for the NPM and destabilizing mode associated with a plastic instability. A scaling analysis close to the instability reveals the surprising finding that the deformation-dynamics of these two mode types follow different scaling laws, both with respect to the distance to the instability strain, and with respect to system size. In particular, we find that
\begin{equation}
\bigg|\frac{d\hat{\Psi}_p}{d\gamma}\bigg|^2 \sim \frac{L^4}{\gamma_c - \gamma}\quad\mbox{and}\quad\bigg|\frac{d\hat{\pi}}{d\gamma}\bigg|^2 \sim \mbox{constant}\,,
\end{equation}
as shown numerically in Figs.~\ref{eigenmode_fig} and \ref{pi_variation_fig}. 

%%%%%%%%%%%%%%%%%%%%%%%%%%%%%%%%%%%%%%%%%%%%%%%%%%%%%%%
%%%%%%%%%%%%%%%%%%%%%%%%%%%%%%%%%%%%%%%%%%%%%%%%%%%%%%

%%%%%%%%%%%%%%%%%%%%%%%%%%%%%%%%%%%%%%%%%%%%%%%%%%%%%%%
%%%%%%%%%%%%%%%%%%%%%%%%%%%%%%%%%%%%%%%%%%%%%%%%%%%%%%

We begin the exploration of the modes' deformation dynamics by constructing the stiffness function
\begin{equation}\label{stiffness_function}
\tilde{\kappa}(\vec{z}) \equiv \frac{{\cal M}:\vec{z}\vec{z}}{\vec{z}\cdot\vec{z}}\,,
\end{equation}
which is a function of a general $N\dbar$-dimensional vector $\vec{z}$, whose single global minimum occurs at $\hat{\Psi}_p$, and $\tilde{\kappa}(\hat{\Psi}_p) = \lambda_p$ is the lowest eigenvalue of ${\cal M}$. Notice that $\tilde{\kappa}(\vec{z})$ as defined above and $\kappa_{\vec{z}}\equiv {\cal M}:\vec{z}\vec{z}$ differ by the normalization that appears in the former but not in the latter. The gradient of $\tilde{\kappa}$ with respect to $\vec{z}$ reads
\begin{equation}
\frac{\partial \tilde{\kappa}}{\partial \vec{z}} = \frac{2}{\vec{z}\cdot\vec{z}}\left({\cal M}\cdot\vec{z} - \frac{\kappa_{\vec{z}}}{\vec{z}\cdot\vec{z}}\vec{z}\right)\,.
\end{equation}
Using the gradient of $\tilde{\kappa}(\vec{z})$ above and the gradient of $b(\vec{z})$ as given by Eq.~(\ref{foo07}), we construct the vector fields 
\begin{eqnarray}
\vec{\Gamma}(\vec{z}) & \equiv & \frac{\vec{z}\cdot\vec{z}}{2}\frac{\partial \tilde{\kappa}}{\partial \vec{z}} = {\cal M}\cdot\vec{z} - \frac{\kappa_{\vec{z}}}{\vec{z}\cdot\vec{z}}\vec{z}\,, \\
\vec{G}(\vec{z}) & \equiv & \frac{\tau^2}{4\kappa^2}\frac{\partial b}{\partial\vec{z}} = {\cal M}\cdot\vec{z} - \frac{\kappa_{\vec{z}}}{\tau_{\vec{z}}}U''':\vec{z}\vec{z}\,. \label{foo27}
\end{eqnarray}
Notice that 
\begin{equation}\label{foo22}
\vec{\Gamma}(\hat{\Psi}_p) = {\cal M}\cdot\hat{\Psi}_p - \lambda_p\hat{\Psi}_p = 0\,,
\end{equation}
and
\begin{equation}\label{foo23}
\vec{G}(\hat{\pi}) = {\cal M}\cdot\hat{\pi} - \frac{\kappa_{\hat{\pi}}}{\tau_{\hat{\pi}}}U''':\hat{\pi}\hat{\pi} = 0\,,
\end{equation}
which motivates the particular definition of $\vec{\Gamma}$ and $\vec{G}$ from the gradients of $\tilde{\kappa}(\vec{z})$ and $b(\vec{z})$ respectively.

The deformation dynamics of the destabilizing mode $\hat{\Psi}_p$ and the NPM $\hat{\pi}$ are derived by requiring that $\hat{\Psi}_p$ and $\hat{\pi}$ remain solutions to Eqs.~(\ref{foo22}) and (\ref{foo23}) under the imposed deformation, namely
\begin{equation}\label{foo16}
\frac{d\vec{\Gamma}}{d\gamma}\bigg|_{\hat{\Psi}_p} = \frac{\partial \vec{\Gamma}}{\partial\gamma}\bigg|_{\hat{\Psi}_p} + \frac{\partial \vec{\Gamma}}{\partial\vec{x}}\bigg|_{\hat{\Psi}_p}\!\!\!\cdot\frac{d\vec{x}}{d\gamma} + \frac{\partial\vec{\Gamma}}{\partial\vec{z}}\bigg|_{\hat{\Psi}_p}\!\!\!\cdot\frac{d\hat{\Psi}_p}{d\gamma} = 0\,,
\end{equation}
and
\begin{equation}\label{foo17}
\frac{d\vec{G}}{d\gamma}\bigg|_{\hat{\pi}}  = \frac{\partial\vec{G}}{\partial\gamma}\bigg|_{\hat{\pi}} + \frac{\partial\vec{G}}{\partial\vec{x}}\bigg|_{\hat{\pi}}\!\!\!\cdot\frac{d\vec{x}}{d\gamma} + \frac{\partial\vec{G}}{\partial\vec{z}}\bigg|_{\hat{\pi}}\!\!\!\cdot\frac{d\hat{\pi}}{d\gamma} = 0\,.
\end{equation}
%\begin{equation}\label{foo16}
% \frac{d}{d\gamma}\frac{\partial \kappa}{\partial\vec{z}}\bigg|_{\hat{\Psi}}\!\!  = \frac{\partial^2 \kappa}{\partial\vec{z}\partial\gamma}\bigg|_{\hat{\Psi}}\!\! + \frac{\partial^2 \kappa}{\partial\vec{z}\partial\vec{x}}\bigg|_{\hat{\Psi}}\!\!\!\cdot\frac{d\vec{x}}{d\gamma} + \frac{\partial^2 \kappa}{\partial\vec{z}\partial\vec{z}}\bigg|_{\hat{\Psi}}\!\!\!\cdot\frac{d\hat{\Psi}}{d\gamma} = 0\,,
% \end{equation}
% and
% \begin{equation}\label{foo17}
% \frac{d}{d\gamma}\frac{\partial b}{\partial\vec{z}}\bigg|_{\hat{\pi}}\!\!  = \frac{\partial^2 b}{\partial\vec{z}\partial\gamma}\bigg|_{\hat{\pi}}\!\! + \frac{\partial^2 b}{\partial\vec{z}\partial\vec{x}}\bigg|_{\hat{\pi}}\!\!\!\cdot\frac{d\vec{x}}{d\gamma} + \frac{\partial^2 b}{\partial\vec{z}\partial\vec{z}}\bigg|_{\hat{\pi}}\!\!\!\cdot\frac{d\hat{\pi}}{d\gamma} = 0\,.
% \end{equation}
Eqs.~(\ref{foo16}) and (\ref{foo17}) can be inverted in favor of $\frac{d\hat{\Psi}_p}{d\gamma}$ and $\frac{d\hat{\pi}}{d\gamma}$ as
\begin{equation}\label{foo18}
\frac{d\hat{\Psi}_p}{d\gamma} = -\bigg(\frac{\partial\vec{\Gamma}}{\partial\vec{z}}\bigg)\!\bigg|_{\hat{\Psi}_p}^{-1}\!\!\cdot \bigg( \frac{\partial\vec{\Gamma}}{\partial\gamma}\bigg|_{\hat{\Psi}_p}\!\! + \frac{\partial\vec{\Gamma}}{\partial\vec{x}}\bigg|_{\hat{\Psi}_p}\!\!\cdot\frac{d\vec{x}}{d\gamma}\bigg) \,,
\end{equation}
and
\begin{equation}\label{foo19}
\frac{d\hat{\pi}}{d\gamma} = -\bigg(\frac{\partial\vec{G}}{\partial\vec{z}} \bigg)\!\bigg|_{\hat{\pi}}^{-1}\cdot\bigg( \frac{\partial\vec{G}}{\partial\gamma}\bigg|_{\hat{\pi}} + \frac{\partial\vec{G}}{\partial\vec{x}}\bigg|_{\hat{\pi}}\!\!\cdot\frac{d\vec{x}}{d\gamma}\bigg)\,.
\end{equation}
%where all partial derivatives should be understood as evaluated at $\hat{\Psi}_p$ (in Eq.~(\ref{foo18})) and $\hat{\pi}$ (in Eq.~(\ref{foo19})). 
%Eqs.~(\ref{foo18}) and (\ref{foo19}) share similar structures, but, as we shall see below, have very different scaling properties with respect to $\gamma_c - \gamma$. 

The analysis of the scaling properties of Eqs.~(\ref{foo18}) and (\ref{foo19}) with respect to $\gamma_c - \gamma$ starts with realizing that $\hat{\Psi}_p$ and $\hat{\pi}$ are zero modes of $\frac{\partial\vec{\Gamma}}{\partial\vec{z}}\big|_{\hat{\Psi}_p}$ and $\frac{\partial\vec{G}}{\partial\vec{z}}\big|_{\hat{\pi}}$, respectively, and therefore $\big(\frac{\partial\vec{\Gamma}}{\partial\vec{z}}\big)\!\big|_{\hat{\Psi}_p}^{-1}$ and $\big(\frac{\partial\vec{G}}{\partial\vec{z}}\big)\!\big|_{\hat{\pi}}^{-1}$ (defined as taken after removal of the zero modes) are regular as $\gamma \to \gamma_c$. Furthermore, the vectors $\frac{\partial\vec{\Gamma}}{\partial\gamma}$ and $\frac{\partial\vec{G}}{\partial\gamma}$ are expected to converge to regular values at plastic instabilities as well. We conclude thus that any singularity that $\frac{d\hat{\Psi}_p}{d\gamma}$ and $\frac{d\hat{\pi}}{d\gamma}$ might possess can only be inherited from the singularity of $\frac{d\vec{x}}{d\gamma}$ (recall that $|\frac{d\vec{x}}{d\gamma}| \sim (\gamma_c - \gamma)^{-1/2}$).

\subsection{Deformation dynamics of destabilizing modes}

Let us focus first on $\frac{d\hat{\Psi}_p}{d\gamma}$ as given by Eq.~(\ref{foo18}); close to instabilities we can approximate $\frac{d\vec{x}}{d\gamma} \simeq -\frac{\nu_p}{\lambda_p}\hat{\Psi}_p$, then
\begin{equation}\label{foo26}
\frac{d\hat{\Psi}_p}{d\gamma} \simeq \frac{\nu_p}{\lambda_p}({\cal M} - \lambda_p{\cal I})^{-1}\cdot (U''':\hat{\Psi}_p\hat{\Psi}_p - \tau_p\hat{\Psi}_p)\,,
\end{equation}
which is singular in terms of $\gamma_c - \gamma$ following the scaling of $\lambda_p \sim \sqrt{\gamma_c - \gamma}$.

%%%%%%%%%%%%%%%%%%%%%%%%%%%%%%%%%%%%%%%%%%%%%%%%%%%%%%%
\begin{figure}[!ht]
\centering
\includegraphics[width = 0.45\textwidth]{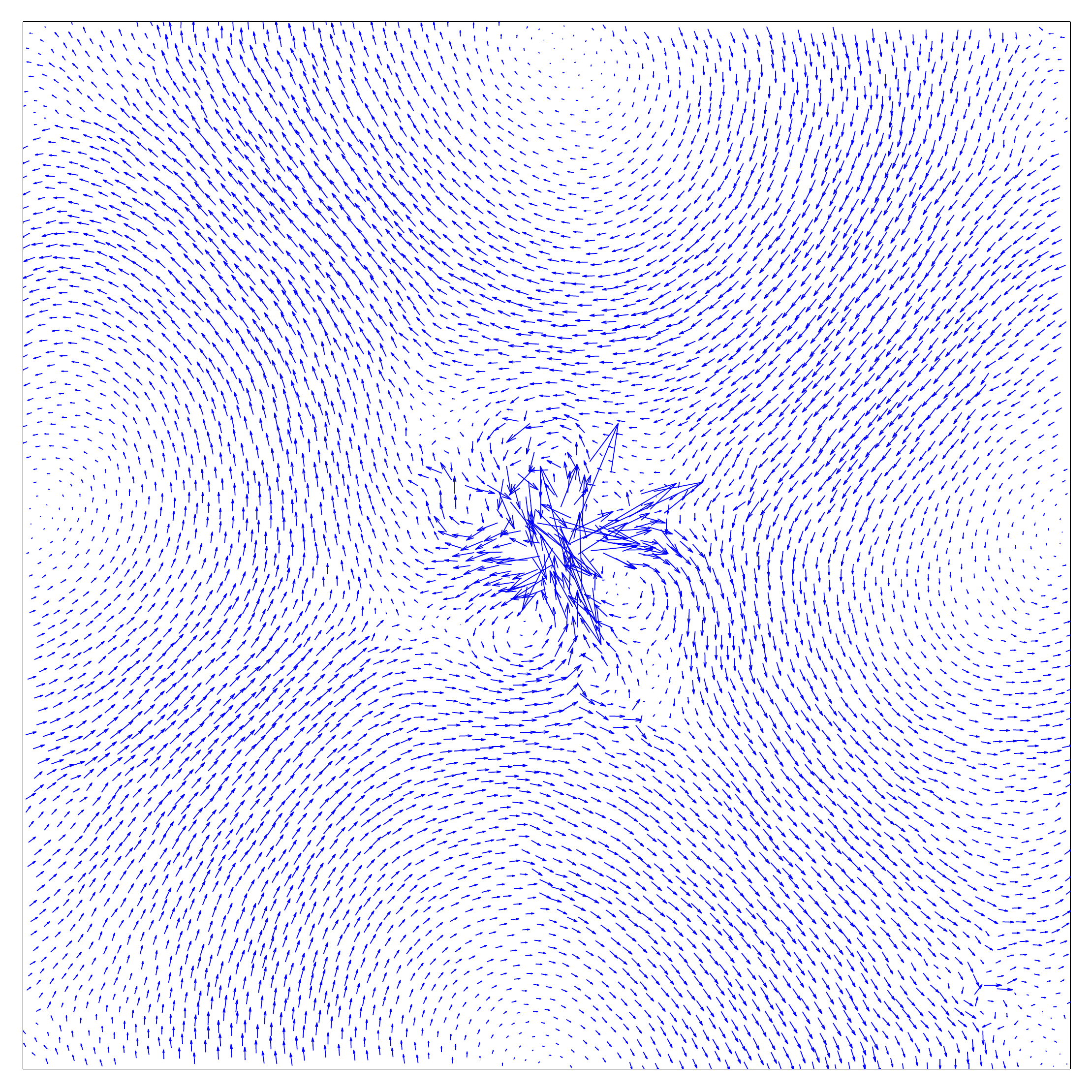}
\caption{\footnotesize  The field $\frac{d\hat{\Psi}_p}{d\gamma}$ calculated at $\gamma_c - \gamma \sim 10^{-14}$ away from a plastic instability in a system of $N=6400$ particles.}
\label{dpsi_dgamma}
\end{figure}
%%%%%%%%%%%%%%%%%%%%%%%%%%%%%%%%%%%%%%%%%%%%%%%%%%%%%%

%In Fig.~\ref{decay_profiles_fig}b it is shown that the vector field $U''':\hat{\Psi}_p\hat{\Psi}_p$ decays at distances $r$ away from the instability core as $r^{-2\dbar}$,
In Fig.~\ref{decay_profiles_fig} it was shown that $\hat{\Psi}$ decays at distances $r$ away from its core as $r^{1-\dbar}$, and $U''':\hat{\Psi}_p\hat{\Psi}_p$ decays as $r^{-2\dbar}$, the former therefore dominates the difference $U''':\hat{\Psi}_p\hat{\Psi}_p - \tau_p\hat{\Psi}_p$ as appears in Eq.~(\ref{foo26}), at large $r$. This difference therefore couples strongly to the lowest-lying eigenmodes of ${\cal M} - \lambda_p{\cal I}$ in Eq.~(\ref{foo26}), which are plane waves with frequencies of order $L^{-1}$. This is further corroborated in Fig.~\ref{dpsi_dgamma}, where we plot the field $\frac{d\hat{\Psi}_p}{d\gamma}$ which displays the same geometry as displayed by the lowest-frequency plane waves of the system. We thus expect 
\begin{equation}\label{foo32}
\bigg|\frac{d\hat{\Psi}_p}{d\gamma}\bigg|^2 \sim \frac{\tau_p^2\nu_p^2L^4}{\lambda_p^2} \sim \frac{\tau_p\nu_p L^4}{\gamma_c-\gamma}\,,
\end{equation}
as found in our numerical simulations, see Fig.~\ref{eigenmode_fig}.

\subsection{Deformation dynamics of NPMs}

We finally turn to analyzing the scaling properties of the equation of motion (\ref{foo19}) for $\frac{d\hat{\pi}}{d\gamma}$. As shown for the case of $\frac{d\hat{\Psi}_p}{d\gamma}$, the only way $\frac{d\hat{\pi}}{d\gamma}$ could be singular in $\gamma_c - \gamma$ is if the RHS of Eq.~(\ref{foo19}) inherits the singularity of $\frac{d\vec{x}}{d\gamma}$, whose norm scales as $(\gamma_c-\gamma)^{-1/2}$. It turns out, however, that $\frac{\partial\vec{G}}{\partial\vec{x}}\big|_{\hat{\pi}}\!\cdot\frac{d\vec{x}}{d\gamma}$ is regular at $\gamma_c$; to see this, we first approximate this contraction close to instabilities as
\begin{widetext}
\begin{equation}\label{foo30}
\frac{\partial\vec{G}}{\partial\vec{x}}\bigg|_{\hat{\pi}}\!\!\cdot\frac{d\vec{x}}{d\gamma} \simeq \frac{\nu_p}{\lambda_p}\left(\frac{ U'''\!\tripleCdot\!\hat{\pi}\hat{\pi}\hat{\Psi}_p}{\tau_{\hat{\pi}}}U'''\!:\!\hat{\pi}\hat{\pi} - U'''\!:\!\hat{\pi}\hat{\Psi}_p + \frac{\kappa_{\hat{z}}}{\tau_{\hat{\pi}}}U''''\tripleCdot\hat{\pi}\hat{\pi}\hat{\Psi}_p - \frac{\kappa_{\hat{\pi}}U''''\quadCdot\hat{\pi}\hat{\pi}\hat{\pi}\hat{\Psi}_p}{\tau_{\hat{\pi}}^2}U''':\hat{\pi}\hat{\pi}\right)\,,
\end{equation}
\end{widetext}
where $U'''' \equiv \frac{\partial^4U}{\partial\vec{x}\partial\vec{x}\partial\vec{x}\partial\vec{x}}$ is the fourth order tensor of derivatives of the potential energy. It is clear that the last two terms on the RHS of the above equation are not singular (they are proportional to $\kappa_{\hat{z}}/\lambda_p$ which approaches unity at the instability strain). We therefore focus for a moment on the first two terms on the RHS of Eq.~(\ref{foo30}); notice that
\begin{equation}\label{foo31}
\sFrac{ U'''\tripleCdot\hat{\pi}\hat{\pi}\hat{\Psi}_p}{\tau_{\hat{\pi}}}U'''\!:\!\hat{\pi}\hat{\pi} - U'''\!:\!\hat{\pi}\hat{\Psi}_p = \sFrac{ U'''\tripleCdot\hat{\pi}\hat{\pi}\vec{\Delta}}{\tau_{\hat{\pi}}}U'''\!:\!\hat{\pi}\hat{\pi} - U'''\!:\!\hat{\pi}\vec{\Delta}\,,
\end{equation}
where we have defined the vector difference $\vec{\Delta} \equiv \hat{\pi} - \hat{\Psi}_p$ between $\hat{\pi}$ and $\hat{\Psi}_p$, and recall that $\tau_{\hat{\pi}} \equiv U'''\tripleCdot\hat{\pi}\hat{\pi}\hat{\pi}$. As the instability strain is approached $\Delta \equiv |\vec{\Delta}| \to 0$, then we can express $\vec{\Delta}$ as the solution to either one of the linear equations:
\begin{eqnarray}
\frac{\partial^2 \tilde{\kappa}}{\partial\vec{z}\partial\vec{z}}\bigg|_{\hat{\Psi}_p}\cdot\vec{\Delta} & = & \frac{\partial\tilde{\kappa}}{\partial\vec{z}}\bigg|_{\hat{\pi}} \label{foo25} \,, \\
\frac{\partial^2 b}{\partial\vec{z}\partial\vec{z}}\bigg|_{\hat{\pi}}\cdot\vec{\Delta} & = & -\frac{\partial b}{\partial\vec{z}}\bigg|_{\hat{\Psi}_p}\,.
\end{eqnarray}
The two above equations are nothing more than the linear expansion of the respective gradients of $\tilde{\kappa}$ and $b$ about their minima at $\hat{\Psi}_p$ and $\hat{\pi}$, respectively. 

%%%%%%%%%%%%%%%%%%%%%%%%%%%%%%%%%%%%%%%%%%%%%%%%%%%%%%%
\begin{figure}[!ht]
\centering
\includegraphics[width = 0.45\textwidth]{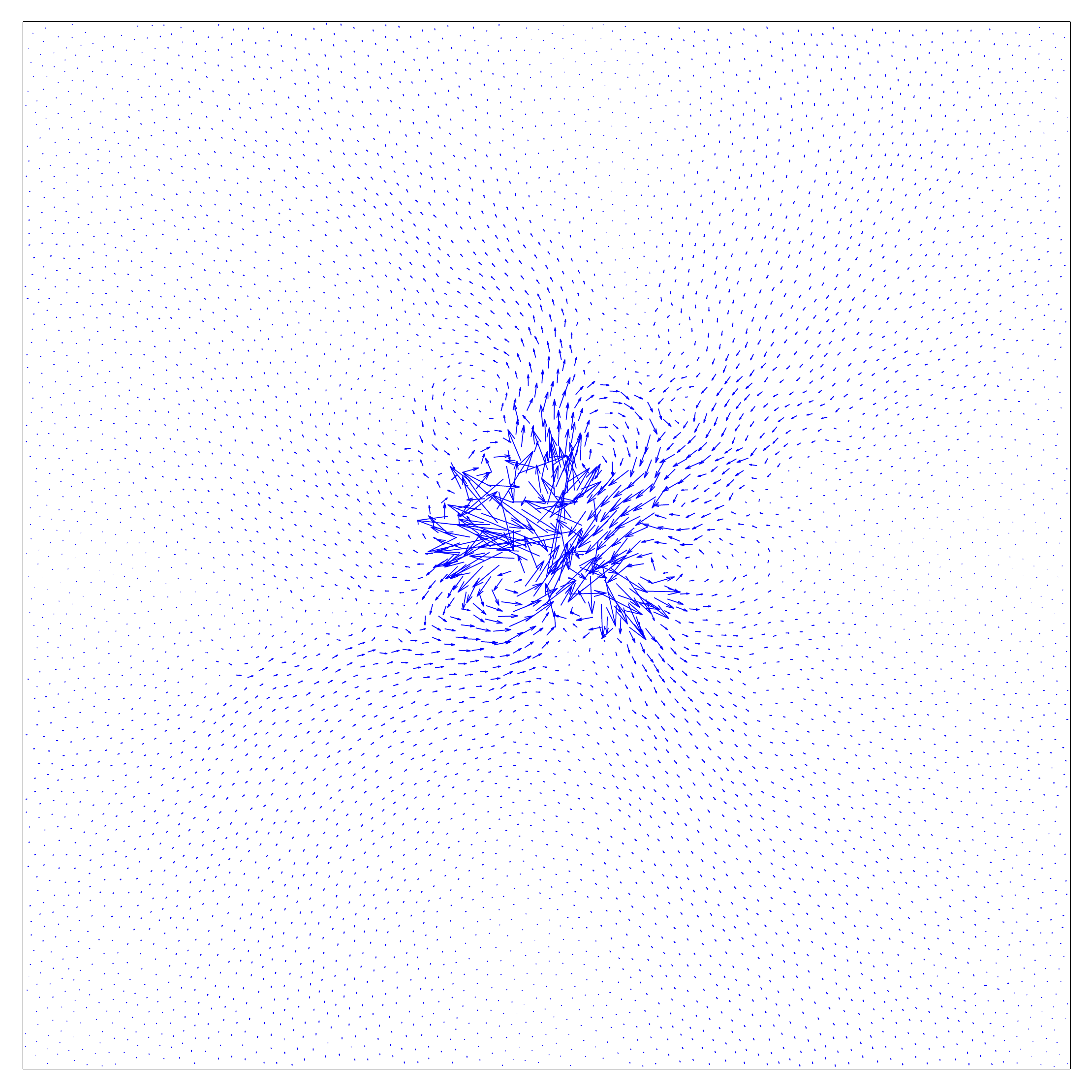}
\caption{\footnotesize  The field $\frac{d\hat{\pi}}{d\gamma}$ calculated at $\gamma_c - \gamma \sim 10^{-14}$ away from a plastic instability in a system of $N=6400$ particles.}
\label{dpi_dgamma}
\end{figure}
%%%%%%%%%%%%%%%%%%%%%%%%%%%%%%%%%%%%%%%%%%%%%%%%%%%%%%

We focus on Eq.~(\ref{foo25}) since it is simpler in structure; taking the partial derivatives, inverting in favor of $\vec{\Delta}$, and using Eq.~(\ref{nonlinear_equation}), we find
\begin{equation}\label{foo29}
\vec{\Delta} \simeq \frac{\kappa_{\hat{\pi}}}{\tau_{\hat{\pi}}}({\cal M} - \lambda_p{\cal I})^{-1}\cdot (U''':\hat{\pi}\hat{\pi} - \tau_{\hat{\pi}}\hat{\pi})\,.
\end{equation}
The above equation explicitly shows that that $\Delta \sim \kappa_{\hat{\pi}} \sim \lambda_p\sim\sqrt{\gamma_c -\gamma}$, which, together with Eqs.~(\ref{foo30}) and (\ref{foo31}) implies that the contraction $\frac{\partial\vec{G}}{\partial\vec{x}}\cdot\frac{d\vec{x}}{d\gamma}$ is regular as $\gamma \to \gamma_c$. This, in turn, implies that $|\frac{d\hat{\pi}}{d\gamma}|$ is regular as well, as discussed above and verified numerically in Fig.~\ref{pi_variation_fig}. Notice that all vectors contracted on the RHS of Eq.~(\ref{foo19}) are of the same order; however in our model glass we find that those that are comprised of contractions with $U''''$ are dominant.

An example of the field $\frac{d\hat{\pi}}{d\gamma}$ is plotted in Fig.~\ref{dpi_dgamma}, calculated at the same instability for which $\frac{d\hat{\Psi}_p}{d\gamma}$ is plotted in Fig.~\ref{dpsi_dgamma}. As opposed to $\frac{d\hat{\Psi}_p}{d\gamma}$ the NPM's variation with strain is a quasi-localized field; this quasi-localization stems from quick spatial decay of the fields $\frac{\partial\vec{G}}{\partial\vec{x}}\big|_{\hat{\pi}}\cdot\frac{d\vec{x}}{d\gamma}$ and $\frac{\partial\vec{G}}{\partial\gamma}\big|_{\hat{\pi}}$ appearing on the RHS of Eq.~(\ref{foo19}). These decay at least as $r^{-2\dbar}$, and therefore do not couple strongly to the low-frequency modes of $\frac{\partial\vec{G}}{\partial\vec{z}}\big|_{\hat{\pi}}$. 

%%%%%%%%%%%%%%%%%%%%%%%%%%%%%%%%%%%%%%%%%%%%%%%%%%%%%%%
\begin{figure}[!ht]
\centering
\includegraphics[width = 0.45\textwidth]{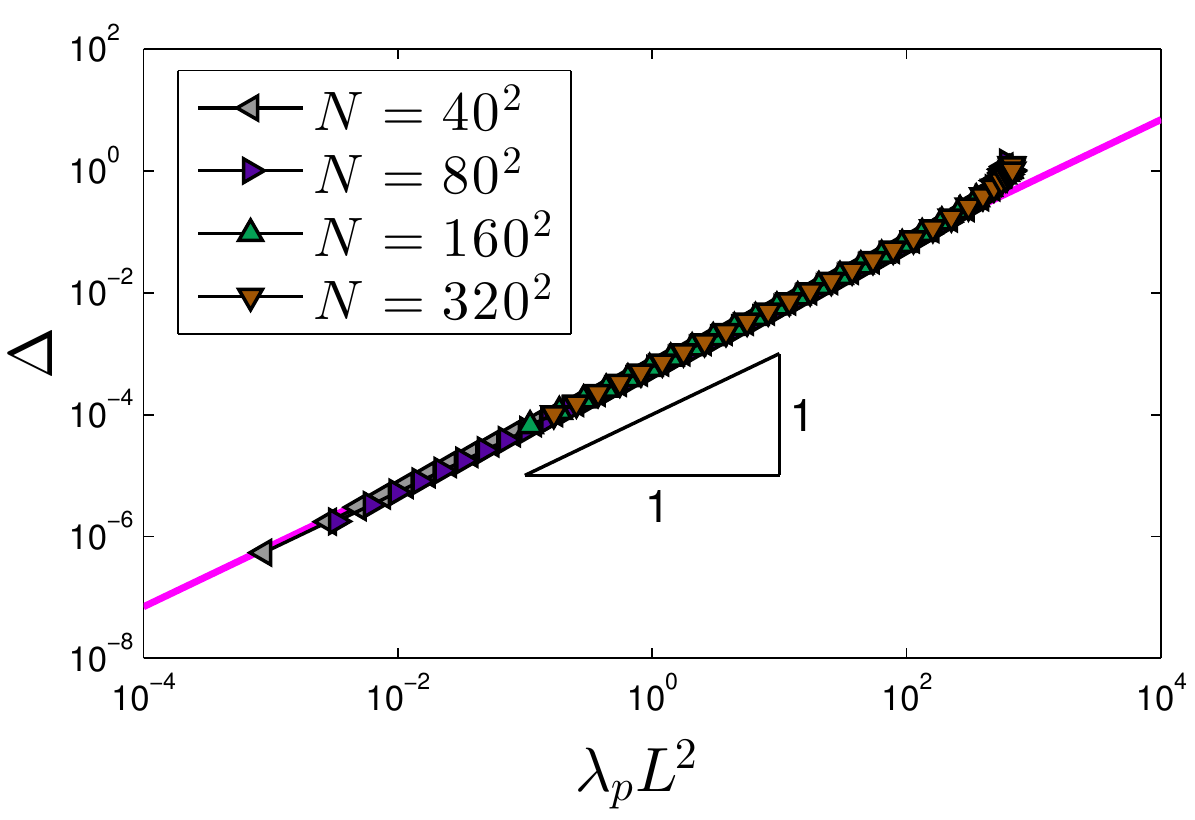}
\caption{\footnotesize Norm $\Delta \equiv |\hat{\pi} - \hat{\Psi}_p|$ of the difference vector between $\hat{\pi}$ and $\hat{\Psi}_p$ vs.~the product $\lambda_pL^2$.}
\label{delta_vs_lambda}
\end{figure}
%%%%%%%%%%%%%%%%%%%%%%%%%%%%%%%%%%%%%%%%%%%%%%%%%%%%%%

Furthermore, as Eq.~(\ref{foo29}) is similar in structure to Eq.~(\ref{foo26}), similar considerations as previously spelled out for $\frac{d\hat{\Psi}_p}{d\gamma}$ apply here as well, and in particular that the far field of $U''':\hat{\pi}\hat{\pi} - \tau_{\hat{\pi}}\hat{\pi}$ is dominated by the slow decay of $\hat{\pi}$ ($\sim r^{1-\dbar}$ see Fig.~\ref{decay_profiles_fig}). It therefore couples strongly to the lowest-lying eigenmodes of ${\cal M} - \lambda_p{\cal I}$, leading to the prediction $\Delta \sim L^2\lambda_p$, shown to hold numerically in Fig.~(\ref{delta_vs_lambda}). By directly comparing Eqs.~(\ref{foo26}) to (\ref{foo29}), we conclude that as $\gamma \to \gamma_c$ 
\begin{equation}
\vec{\Delta} \equiv \hat{\pi} - \hat{\Psi} \simeq \frac{\lambda_p^2}{\tau_p\nu_p}\frac{d\hat{\Psi}_p}{d\gamma}\,,
\end{equation}
which means that $\vec{\Delta}$ has the structure of the lowest-frequency plane-wave, as can be seen in Fig.~\ref{dpsi_dgamma}.

\subsection{Predictiveness of NPMs}
\label{predictiveness}
In the previous two subsections we have shown that there is a dramatic difference between the deformation-dynamics of destabilizing modes compared to that of NPMs. Although their associated stiffnesses share the same scaling with $\gamma_c - \gamma$ close to instabilities (see Eqs.~(\ref{foo01}) and (\ref{kappa})), the two types of modes exhibit different scaling laws in their variation rate as an instability is approached, and in particular 
\[
\bigg| \frac{d\hat{\Psi}_p}{d\gamma} \bigg|^2 \sim \frac{L^4}{\gamma_c - \gamma}\,, \quad \mbox{but} \quad \bigg| \frac{d\hat{\pi}}{d\gamma} \bigg|^2 \sim \mbox{constant}\,.
\]

To what degree do destabilizing modes and NPMs indicate their common final form away from the instability strain? This can be quantified by considering the differences $1 - \hat{\pi}(\gamma)\cdot\hat{\pi}(\gamma_c)$ and $1-\hat{\Psi}_p(\gamma)\cdot\hat{\Psi}_p(\gamma_c)$ for the NPM and the destabilizing mode cases, respectively. The former can be easily estimated by Taylor expanding $\hat{\pi}(\gamma)$ around $\gamma_c$ (which is possible due to its regularity), leading to the prediction
\begin{equation}\label{foo12}
1 - \hat{\pi}(\gamma)\cdot\hat{\pi}(\gamma_c) \sim (\gamma_c - \gamma)^2\,, 
\end{equation}
where we have used that $\frac{d\hat{\pi}}{d\gamma}\cdot\hat{\pi} = 0$. 

%%%%%%%%%%%%%%%%%%%%%%%%%%%%%%%%%%%%%%%%%%%%%%%%%%%%%%%
\begin{figure}[!ht]
\centering
\includegraphics[width = 0.48\textwidth]{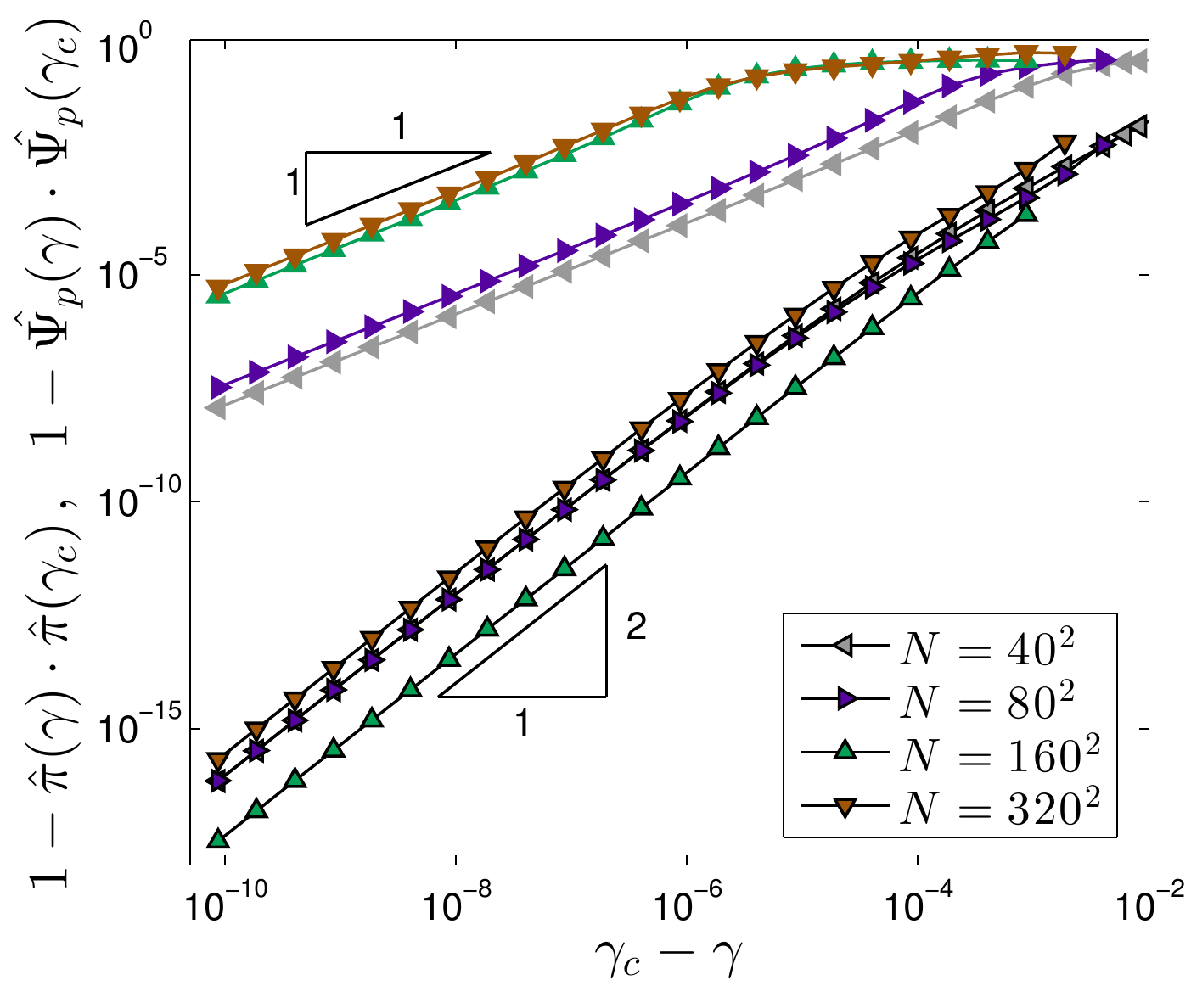}
\caption{\footnotesize $1-\hat{\pi}(\gamma)\cdot\hat{\pi}(\gamma_c)$ (outlined symbols) and $1-\hat{\Psi}_p(\gamma)\cdot\hat{\Psi}_p(\gamma_c)$ (solid symbols) \emph{vs}.~$\gamma_c - \gamma$. NPMs converge must faster \emph{scaling-wise} to their final form at instabilities compared to destabilizing modes, and are therefore better predictors of imminent plastic instabilities.}
\label{convergence_fig}
\end{figure}
%%%%%%%%%%%%%%%%%%%%%%%%%%%%%%%%%%%%%%%%%%%%%%%%%%%%%%

The destabilizing mode case is slightly more subtle due to the singularity in its derivative as seen in Eq.~(\ref{foo26}). However, since the said singularity is integrable, we can define
\[
\delta\vec{\Psi} \equiv \hat{\Psi}_p(\gamma_c) - \hat{\Psi}_p(\gamma) = \int\limits_{\gamma_c}^\gamma\! \frac{d\hat{\Psi}_p}{d\gamma}\bigg|_{\tilde{\gamma}}\! d\tilde{\gamma}\,,
\]
with the norm $|\delta\vec{\Psi}| \sim L^2\sqrt{\gamma_c - \gamma}$ following Eq.~(\ref{foo32}). Notice that $|\delta\vec{\Psi}|^2 = 2 - 2\hat{\Psi}_p(\gamma_c)\cdot\hat{\Psi}_p(\gamma)$, therefore we predict
\begin{equation}\label{foo13}
1-\hat{\Psi}_p(\gamma_c)\cdot\hat{\Psi}_p(\gamma) \sim L^4(\gamma_c - \gamma)\,.
\end{equation}

The scaling laws Eqs.~(\ref{foo12}) and (\ref{foo13}) are verified numerically in Fig.~\ref{convergence_fig}. They further explain the observation that away from instabilities the overlaps $1-\hat{\pi}\cdot\hat{\Psi}_p \sim L^4(\gamma_c - \gamma)$, as seen in Fig.~\ref{overlap_pi_psi}: since NPMs converge very quickly to their final forms at the instability, $\hat{\pi}\cdot\hat{\Psi}_p$ is bounded by the convergence rate of the destabilizing mode, as given by Eq.~(\ref{foo13}). 

Besides the difference in convergence rates between the two mode types as seen in Fig.~\ref{convergence_fig}, perhaps the most striking feature of this data is the typical value measured for $\hat{\pi}(\gamma)\cdot\hat{\pi}(\gamma_c)$ when the NPMs are first detected, at strain scales on the order of $10^{-3}$ away from the instability. At these strains the overlaps with the final form of the NPMs agree up to a few tenths of a percent, indicating that once detected, NPMs are nearly perfect indicators of the locus and geometry of imminent plastic instabilities. 

\section{Summary and outlook}
\label{summary}

We have carried out a comparative theoretical and numerical analysis of the deformation dynamics of nonlinear plastic modes and destabilizing eigenmodes upon approaching plastic instabilities. We have found that although the stiffnesses associated with these two mode types follow the same scaling with strain, the modes themselves vary with vastly different rates as instabilities are approached. Not only do NPMs not suffer from hybridizations with low frequency normal modes as destabilizing modes do, but their variation rate is regular upon approaching plastic instabilities, in stark contrast with the singular variation rate of destabilizing modes. These results add substantial support to the usefulness of NPMs as robust plasticity predictors, and to the role NPMs' spatial distribution may play as a state variable that controls the rate of plastic deformation in glasses subjected to external loading.

The picture that emerges from our study is that the system size and strain dependence in the deformation dynamics of destabilizing mode stems from the dehybridization process that continues to take place all the way up to the instability strain. We find that close to plastic instabilities the destabilizing mode can be obtained by adding a plane-wave-like mode with an amplitude proportional to $L^2\sqrt{\gamma_c - \gamma}$ to the NPM. This interpretation suggests that the most relevant objects to plastic flow in disordered solids are NPMs, and that research efforts should be focused on studying their statistics and dynamics. 

Our analysis reveals that a NPM $\hat{\pi}$ is characterized by three key physical parameters: the stiffness $\kappa_{\hat{\pi}}$, the asymmetry $\tau_{\hat{\pi}}$, and the shear-force coupling $\nu_{\hat{\pi}}$. A local instability strain can be defined using these parameters as $\delta\gamma_{\hat{\pi}} \equiv \frac{\kappa_{\hat{\pi}}^2}{2\nu_{\hat{\pi}}\tau_{\hat{\pi}}}$, following Eq.~(\ref{kappa}). While similar modes are expected to form local minima of $\delta\gamma_{\hat{z}}$ (written as a function of a general $N\dbar$-dimensional displacement direction $\hat{z}$) and of the barrier function $b(\hat{z})$ reintroduced in this work, the deformation dynamics as presented in this work do not strictly speaking hold for minima of $\delta\gamma_{\hat{z}}$. One can nevertheless use $\delta\gamma_{\hat{\pi}}$ (i.e.~evaluated at NPMs $\hat{\pi}$ calculated using the barrier function) as an indicator of the proximity of an individual NPM to its particular plastic instability strain. 

One important question we leave for future research is whether correlations exist between the amount of energy dissipated in an elementary shear transformation, and the parameters $\tau_{\hat{\pi}}$ and $\nu_{\hat{\pi}}$ associated with the NPM that destabilized. In other words, can the post-instability consequences be predicted based on pre-instability information? Considering e.g.~the observed variance between samples of the prefactors of the scaling $\kappa_{\hat{\pi}} \sim \sqrt{\gamma_c - \gamma}$, and of the variation rates $\frac{d\hat{\pi}}{d\gamma}$, it is possible that besides predicting the strain at which an NPM would destabilize, this information might be indicative of post-instability mechanics. 

In this work we did not touch upon the important task of a-priori detecting of the entire field of NPMs of a solid. The usefulness of the NPM framework clearly hinges on the availability of computational methods that are able to robustly detect this field and monitor its statistics and dynamics. Such methods are currently under development, and are left for future studies. 

%Preliminary results indicate that this is a feasible task due to two characteristics of the landscape of the barrier function. Firstly, we find a correlation between the value of $b(\hat{z})$ at local minima, and the volume in direction space of their associated basin. In particular, lower-lying minima are associated to larger basins. This facilitates the computational task of finding the most relevant local minima of $b(\vec{z})$ -- the lowest lying ones. Secondly, the landscape of $b(\hat{z})$ is smoother relative to other landscapes whose minima provide information about imminent plastic instabilities e.g.~the function $|\frac{\partial U}{\partial\vec{x}}|^2$ \cite{}. This statement is supported by e.g.~the observation that the gradient of $b(\hat{z})$ is bounded

\acknowledgements
We acknowledge Luka Gartner for providing analysis codes. We warmly thank Gustavo D\"uring and Eran Bouchbinder for fruitful discussions. 

\appendix
\section{Tensoric notation convention}
In this work we omit particle indices with the goal of improving the clarity and readability of the text. We denote $N\dbar$-dimensional vectors as $\vec{v}$, each component pertains to some particle index $i$ and a particular Cartesian spatial component. Tensors defined as derivatives with respect to coordinates $\vec{x}$ or the displacements $\vec{z}$ are denoted e.g.~$\frac{\partial^3U}{\partial\vec{x}\partial\vec{x}\partial\vec{x}}$, which should be understood as $\frac{\partial^3U}{\partial\vec{x}_i\partial\vec{x}_j\partial\vec{x}_k}$ with $i,j,k$ denoting particle indices. Single, double, triple and quadruple contractions are denoted by $\cdot$, $:$, $\tripleCdot$, and $\quadCdot$, respectively. For example, the RHS of Eq.~(\ref{foo27}) ${\cal M}\cdot\vec{z} - \frac{\kappa_{\vec{z}}}{\tau_{\vec{z}}}U''':\vec{z}\vec{z}$ should be interpreted as
\begin{equation}
 {\cal M}_{ij}\cdot\vec{z}_j - \frac{\kappa_{\vec{z}}}{\tau_{\vec{z}}}\frac{\partial^3U}{\partial\vec{x}_i\partial\vec{x}_j\partial\vec{x}_k}:\vec{z}_j\vec{z}_k\,,
\end{equation}
where repeated indices should be understood as summed over.

\section{Models and numerical methods}
We employ a 50:50 binary mixture of `large' and `small' particles of equal mass $m$ in two dimensions, interacting via radially-symmetric purely repulsive inverse power-law pairwise potentials, that follow
\begin{equation}
\varphi_{\mbox{\tiny IPL}}(r_{ij}) = \left\{ \begin{array}{ccc}\varepsilon\left[ \left( \sFrac{a_{ij}}{r_{ij}} \right)^n + \sum\limits_{\ell=0}^q c_{2\ell}\left(\sFrac{r_{ij}}{a_{ij}}\right)^{2\ell}\right]&,&\sFrac{r_{ij}}{a_{ij}}\le x_c\\0&,&\sFrac{r_{ij}}{a_{ij}}> x_c\end{array} \right.,
\end{equation}
where $r_{ij}$ is the distance between the $i^{\mbox{\tiny th}}$ and $j^{\mbox{\tiny th}}$ particles, $\varepsilon$ is an energy scale, and $x_c$ is the dimensionless distance for which $\varphi_{\mbox{\tiny IPL}}$ vanishes continuously up to $q$ derivatives. Distances are measured in terms of the interaction lengthscale $a$ between two `small' particles, and the rest are chosen to be $a_{ij} = 1.18a$ for one `small' and one `large' particle, and $a_{ij} = 1.4a$ for two `large' particles. The coefficients $c_{2\ell}$ are given by
\begin{equation}
c_{2\ell} = \frac{(-1)^{\ell+1}}{(2q-2\ell)!!(2\ell)!!}\frac{(n+2q)!!}{(n-2)!!(n+2\ell)}x_c^{-(n+2\ell)}\,.
\end{equation}
We chose the parameters $x_c = 1.48, n=10$, and $q=3$. The density was set to be $N/V = 0.86a^{-2}$. This model undergoes a computer-glass-transition around the temperature $T_g\approx 0.5\varepsilon/k_B$. Solids were created by a fast quench from the melt to a target temperature $T\ll T_g$, followed by an energy minimization using a standard nonlinear conjugate gradient algorithm. Systems were deformed by imposing simple shear, meaning that the coordinates $x_i,y_i$ of each particle were displaced according to 
\begin{eqnarray}
x_i & \to & x_i + \delta\gamma y_i\,, \\
y_i & \to & y_i\,,
\end{eqnarray}
where $\delta\gamma$ is the strain increment, chosen to be smaller than $10^{-3}$. 128-bit numerics were employed, which enabled us to approach  instabilities up to $\gamma_c - \gamma \approx 10^{-14}$. 

Once each system was brought as closely as possible to the firstly encountered plastic instability, the lowest eigenmode of ${\cal M}$ was calculated by minimizing the stiffness function $\tilde{\kappa}(\vec{z})$ as given by Eq.~(\ref{stiffness_function}) over directions $\vec{z}$. The minimization was carried out via a standard nonlinear conjugate gradient algorithm, while the norm of $\vec{z}$ was monitored and maintained during the minimization. $\tilde{\kappa}$ has a single minimum at the lowest eigenmode $\hat{\Psi}_p$ of ${\cal M}$, which is uncovered upon convergence of the minimizer. This allows us to start this minimization with any random initial conditions $\hat{z}_{\mbox{\tiny ini}}$; the minimization is guaranteed to terminate with $\hat{\Psi}_p$. 

Once calculated, the eigenmode $\hat{\Psi}_p$ found close to an instability strain $\gamma_c$ is then used for all subsequent calculations of nonlinear plastic modes away from the instability strain. This is done at each strain by minimizing the barrier function $b(\vec{z})$ as given by Eq.~(\ref{barrier_function}), with the eigenmode $\hat{\Psi}_p\big|_{\gamma \to \gamma_c}$ as the initial conditions for the minimization. The same minimization code for $\tilde{\kappa}(\vec{z})$ is used for minimizing $b(\vec{z})$. 

Derivatives with respect to strain of eigenmodes $\hat{\Psi}_p$ and NPMs were calculated by finite differences. The results were validated close to the instability strains by directly solving Eqs.~(\ref{foo18}) and (\ref{foo19}). 

\vspace{1cm}

\section{Double contractions with the third-order tensor $\frac{\partial^3U}{\partial\vec{x}\partial\vec{x}\partial\vec{x}}$}

In this Appendix we motivate Eq.~(\ref{foo08}) of the main text, and in particular we show that the double contraction of $U''' \equiv \frac{\partial^3U}{\partial\vec{x}\partial\vec{x}\partial\vec{x}}$ with a field characterized by some spatial variation is expected to scale as the square of the gradient of that field, for the case of pairwise central-force potentials. 

Assuming the potential energy is written as $U = \sum_{i<j}\varphi_{ij}$, with $\varphi$ the pairwise central potential, the tensor of interest is
\begin{widetext}
\begin{equation}
\sFrac{\partial^3 U}{\partial\vec{x}_\ell\partial\vec{x}_m\partial\vec{x}_n} = \sum_{i<j} \varphi'''_{ij}\sFrac{\partial r_{ij}}{\partial \vec{x}_\ell}\sFrac{\partial r_{ij}}{\partial \vec{x}_m}\sFrac{\partial r_{ij}}{\partial \vec{x}_n} + \sum_{i<j}\varphi''_{ij}\left(\sFrac{\partial^2r_{ij}}{\partial\vec{x}_\ell\partial\vec{x}_m}\sFrac{\partial r_{ij}}{\partial\vec{x}_n} + \sFrac{\partial^2r_{ij}}{\partial\vec{x}_\ell\partial\vec{x}_n}\sFrac{\partial r_{ij}}{\partial\vec{x}_m} + \sFrac{\partial^2r_{ij}}{\partial\vec{x}_m\partial\vec{x}_n}\sFrac{\partial r_{ij}}{\partial\vec{x}_\ell}\right) +  \sum_{i<j} \varphi'_{ij} \sFrac{\partial ^3r_{ij}}{\partial\vec{x}_\ell\partial\vec{x}_m\partial\vec{x}_n} \,,
\end{equation}
\end{widetext}
with $\varphi_{ij}' \equiv \frac{\partial\varphi}{\partial r_{ij}}$ etc., and $r_{ij}\equiv \sqrt{\vec{x}_{ij}\cdot\vec{x}_{ij}}$ is the pairwise distance between particles $i$ and $j$, and $\vec{x}_{ij} \equiv \vec{x}_j - \vec{x}_i$. A direct calculation shows that 
\begin{eqnarray}
\sFrac{\partial r_{ij}}{\partial \vec{x}_\ell}\cdot\vec{v}_\ell & \sim & |\vec{v}_{ij}| \,, \nonumber \\
\sFrac{\partial^2r_{ij}}{\partial\vec{x}_\ell\partial\vec{x}_m}\sFrac{\partial r_{ij}}{\partial\vec{x}_n}:\vec{v}_m\vec{v}_n & \sim &  |\vec{v}_{ij}|^2 \,, \nonumber \\
\sFrac{\partial^2r_{ij}}{\partial\vec{x}_\ell\partial\vec{x}_m}:\vec{v}_\ell\vec{v}_m & \sim &  |\vec{v}_{ij}|^2 \,, \nonumber \\
\sFrac{\partial^3r_{ij}}{\partial\vec{x}_\ell\partial\vec{x}_m\partial\vec{x}_n}:\vec{v}_m\vec{v}_n & \sim &  |\vec{v}_{ij}|^2 \,, \nonumber 
\end{eqnarray}
If the interaction $\varphi$ is short-ranged then the dominant contribution to the contraction $U''':\vec{v}\vec{v}$ comes from the first coordination shells. For those pairs, $|\vec{v}_{ij}| \sim |\nabla \vec{v}|$, and therefore $|U''':\vec{v}\vec{v}| \sim |\nabla \vec{v}|^2$, as expressed by Eq.~(\ref{foo08}). 

\vspace{3.5cm}

%Next notice that for a pair $i,j$ and an arbitrary vector $\vec{v}$
%\begin{eqnarray}
%\sFrac{\partial r_{ij}}{\partial \vec{x}_\ell}\cdot\vec{v}_\ell & = & \sFrac{\vec{v}_{ij}\cdot\vec{x}_{ij}}{r_{ij}} \sim |\vec{v}_{ij}|\,, \nonumber \\
%\sFrac{\partial^2r_{ij}}{\partial \vec{x}_\ell\partial \vec{x}_m}:\vec{v}_\ell\vec{v}_m & = & \sFrac{\vec{v}_{ij}\cdot\vec{v}_{ij}}{r_{ij}} - \sFrac{(\vec{x}_{ij}\cdot\vec{v}_{ij})^2}{r_{ij}^3}\sim |\vec{v}_{ij}|^2\,, \nonumber \\
%\sFrac{\partial^2r_{ij}}{\partial \vec{x}_\ell\partial \vec{x}_m}\sFrac{\partial r_{ij}}{\partial\vec{x}_n}:\vec{v}_m\vec{v}_n & = & \sFrac{\vec{v}_{ij}\cdot\vec{x}_{ij}}{r_{ij}}\big(\sFrac{\vec{v}_{ij}}{r_{ij}} - \sFrac{\vec{v}_{ij}\cdot\vec{x}_{ij}\vec{x}_{ij}}{r_{ij}^3}  \big)\sim |\vec{v}_{ij}|^2 \nonumber \\
%\end{eqnarray}s

\bibliography{references_lerner}

\end{document}